%% file: Lignieres_ar.tex
\documentclass{aa} 

\usepackage{graphicx}
\usepackage{txfonts}
\usepackage{mathrsfs}
\input{commands.tex}
\newcommand{\ri}{r_{\rm i}}
\newcommand{\ro}{r_{\rm o}}

\newcommand{\Lop}{\ensuremath{{\cal L}}}
\newcommand{\LopL}{\ensuremath{\Lop_\mathrm{L}}}
\newcommand{\vb}[1]{\ensuremath{\boldsymbol{#1}}}
\newcommand{\vxi}{\ensuremath{\boldsymbol{\xi}}}

\newcommand{\er}{\ensuremath{\boldsymbol{e}_r}}
\newcommand{\etheta}{\ensuremath{\boldsymbol{e}_\theta}}
\newcommand{\ephi}{\ensuremath{\boldsymbol{e}_\phi}}
\begin{document} 
   \title{Perturbative analysis of the effect of a magnetic field on gravito-inertial modes}
   \author{F. Lignières \inst{1} \and J. Ballot \inst{1} \and S. Deheuvels \inst{1} \and M. Galoy \inst{1}}
   
   \date{Received October 12, 2023; Accepted November 22, 2023}
   
   \institute{IRAP, Université de Toulouse, CNRS, CNES, UPS, 31400 Toulouse, France\\
              \email{francois.lignieres@irap.omp.eu}}
              
% \abstract{}{}{}{}{} 
% 5 {} token are mandatory
 
  \abstract
  {Magnetic fields have been measured recently in the core of red giant stars thanks to their effects on stellar oscillation frequencies. The search for magnetic signatures in pulsating stars, such as $\gamma$ Doradus or Slowly Pulsation B stars, requires to adapt the formalism developed for the slowly rotating red giants to rapidly rotating stars.}
   {We perform a theoretical analysis of the effects of an arbitrary  magnetic field on high radial order gravity and Rossby modes in a rapidly rotating star.}
   {The magnetic effects are
treated as a perturbation. %jb %perturbatively. 
For high radial order modes, the contribution of the radial component of the magnetic field is likely to dominate over the azimuthal and latitudinal components. The rotation is taken into account through the traditional approximation of rotation.}
   {General expressions of the frequency shift induced by an arbitrary radial magnetic field are derived. Approximate analytical forms are obtained in the high-order high-spin-parameter limits for the modes most frequently observed in $\gamma$ Dor stars. We propose simple methods to detect seismic magnetic signatures and measure possible %?jb %eventual 
   magnetic fields in such stars.}
   {These methods offer new possibilities to look for internal magnetic fields in future observations, such as the ones of the PLATO mission, or to revisit existing \emph{Kepler} or TESS data.} 
   \maketitle
%
%-------------------------------------------------------------------

%\onecolumn

\section{Introduction}

The transport of angular momentum and chemical elements in stellar radiative zones is a key issue of modern stellar evolution theory.
Purely hydrodynamical models fail to account for all the observational constraints,
in particular they are not efficient enough to reproduce the core rotation rates of red giant or intermediate-mass stars measured by seismology \citep{MG13,CE13,OM19}.
Models involving magnetic fields are a promising way of solving this angular momentum transport problem \citep{GM98, FP19, GJ22}.
However, observational constraints on the strength and topology of radiative zone magnetic fields are rare,  limiting our ability to test these models.

Spectropolarimetry enables
magnetic field measurements at the surface of stars
with radiative enveloppe. It was found that  5-10\%  of intermediate-mass and massive main-sequence stars
have detectable magnetic fields. The vast majority of them are strong (> 100 Gauss), stable in time and with a simple
large scale topology \citep{AW07,Donati09,WN16}, whereas very weak ($\sim$ 1 Gauss) fields have been detected on a few very bright intermediate-mass stars \citep{L14,BP16}. 
Asteroseismology has recently provided constraints on internal magnetic fields.
First the absence of the dipolar mixed gravito-acoustic modes in the oscillation spectrum
of a fraction ($\sim$ 20\%) of red giant stars has been interpreted as due to strong magnetic fields in their radiative core \citep{FC15,SC16}.
Accordingly, if magnetic fields exceeds a certain limit, gravity waves transform into short-wavelength Alfvèn waves that soon get dissipated \citep{LV17,RF23}.
It has been nevertheless argued that this interpretation
is incompatible with the seismic properties of stars with weakly depressed dipolar mixed modes \citep{MB17}.
Second a clear signature of magnetic fields, the asymmetry of the $\ell$=1 triplet, has been discovered in the oscillation spectrum of 13 red giant stars.
This data together with the theoretical description of the effects
of an arbitrary magnetic field on these oscillation modes enabled to measure the strength of the radial
component of the magnetic field in the vicinity of the H-burning shell of these stars
and  to constrain their topology \citep{LD22, LD23}.
A related seismic signature, the deviation from the uniform period spacing of gravity modes, led to the detection of core magnetic fields in 11 more red giant stars \citep{DL23}.

Beyond red giants, seismic signatures of magnetic fields can of course be searched in other pulsating stars.
$\gamma$ Doradus stars are good candidates  because (i)
they oscillate in the low-frequency range, which is most affected by magnetic fields (ii) their modes are gravity and Rossby modes which are sensitive to the radiative layers just outside the convective
core where a magnetic field is most probably generated by a convective dynamo, (iii) these gravity and Rossby modes are identified
in more than 600 $\gamma$ Dor stars  \citep{LV20} observed by \emph{Kepler} \citep{Borucki10} and in about 60 $\gamma$ Dor stars  \citep{Garcia22_gDor} observed by TESS \citep[Transiting Exoplanet Survey Satellite,][]{Ricker15_TESS}.
However, the two seismic diagnostics available to search for magnetic field in the slowly rotating red giants, namely
the asymmetry of the $\ell$=1 triplet and the deviation from the g-mode uniform period spacing,
are not relevant for rapidly rotating stars such as $\gamma$ Doradus star. A theoretical analysis of the effects of an arbitrary magnetic field
in a rapidly rotating star is therefore necessary to search for and interpret potential magnetic seismic signatures in $\gamma$ Dors.

The effects of a magnetic field on oscillation modes were first considered for acoustic modes in \citet{UO89,GT90}.
Both the magnetic field and the rotation
were assumed small enough to be treated perturbatively. This approach has been later applied to high-order gravity modes, including mixed gravito-acoustic modes. In these studies, the poloidal magnetic field has been assumed to take the simple form of a dipolar field either aligned \citep{HZ05,GL20,BP21,MB21} or inclined \citep{L21}
with respect to the rotation axis.  In \cite{LD22}, we extended to an arbitrary field the perturbative analysis of magnetic and rotation effects on gravity modes. This extension is significant as we do not want seismic measurements of magnetic field to depend on a priori assumption on the magnetic field topology. 

In $\gamma$ Doradus stars, the rotation can not be treated with a perturbative method as the Coriolis force significantly affects the gravity modes (hereafter the g modes) and induces new modes such as the Rossby modes (hereafter the r modes). A well-known approximation, the Traditional Approximation of Rotation (TAR), allows to take Coriolis
effects into account while keeping the eigenvalue problem separable in the spherical coordinates.
Although the TAR has its limitations \citep{F87,GS05,BL12, OS17, OL20},
it is most often accurate and has been successful in
identifying oscillation modes
and measuring core rotation rates in $\gamma$ Doradus stars
\citep{BM15,VT15,VT16,CB18,LV19,LV20}.
The magnetic effects on TAR gravito-inertial modes have been already studied in the case of dipolar poloidal magnetic fields either aligned \citep{PM19} or inclined \citep{PM20}
with respect to the rotation axis. In both studies, frequency shifts induced by the dipolar magnetic field have been computed
for one particular stellar model.
In the present paper,  we extend these works (i) by considering an arbitrary magnetic field and (ii) by deriving approximate analytical formulae relating
the magnetic frequency shifts to a weighted average of the internal magnetic field, the rotation rate, and stellar structure properties for the modes
that are most frequently identified
in $\gamma$ Doradus stars.
We then proprose simple methods to search for magnetic seismic signatures and potentially measure  magnetic fields in the frequency spectra of $\gamma$ Dor or SPB stars.

The paper is organized as follow: the formula of the frequency shift induced by an arbitrary magnetic field on high-order gravito-inertial modes is derived in Sect.~\ref{first}.
In Sect.~\ref{second}, properties of the magnetic frequency shift are studied for the modes observed in $\gamma$ Dor: approximate analytical formulas are derived and tested (Sect.~\ref{sub1}), the magnetic shift of the different modes are compared to help detect magnetic signature in observed spectra (Sect.~\ref{sub3}), methods to detect magnetic signatures are tested on synthetic spectra  (Sect.~\ref{sub2}). In Sect.~\ref{disc}, the results are summarized and discussed.

\section{Derivation of the magnetic shift} 
\label{first}

Time-harmonic $\propto \exp(i \omega t)$ small perturbations of a magnetic and uniformly rotating star are solutions of the following eigenvalue problem:
\begin{equation}
\label{start}
\omega^2 \vxi = 2 i \omega \vec{\Omega} \wedge \vxi + \Lop_0(\vxi) + \LopL(\vxi)
\end{equation}
where $\vxi$ is the displacement vector, $2 i \omega \vec{\Omega} \wedge \vxi$ is the Coriolis force term,  $\Lop_0$ is a linear operator 
\begin{equation}
\Lop_0(\vxi) = \rho^{-1}\vb{\nabla} p' + \vb{\nabla}\psi' +\rho^{-1}\vb{\nabla}\psi\rho' ,
\end{equation}
that has the same form as in a non-rotating non-magnetic star
with $\rho$ and $\psi$ the equilibrium density and gravitational potential, $p'$, $\rho'$ and $\psi'$ the perturbations to pressure, density and gravitational potential,
and $\LopL(\vxi)$ is the linear operator associated with the Lorentz force,
\begin{equation}
\LopL(\vxi)
    =  -\frac{1}{\rho\mu_0} [(\vb{\nabla} \times \vb{B}') \times \vb{B} + (\vb{\nabla} \times \vb{B}) \times \vb{B}' ] - \frac{\vb{\nabla}\cdot(\rho\vxi)}{\rho^2\mu_0} [(\vb{\nabla} \times \vb{B}) \times \vb{B} ]
\end{equation}
with $\vb B$ the star magnetic field, $\vb{B}' = \vb{\nabla} \times (\vxi \times \vb{B})$ the Eulerian perturbation to $\vb{B}$, and $\mu_0$ the magnetic permeability (e.g. Li et al. \citeyear{LD22}).

We perform a first order perturbative analysis of the effects of a magnetic field assuming $\|\LopL(\vxi)\| \ll \|\Lop_0(\vxi) + 2 i \omega \vec{\Omega} \wedge \vxi\|$. 
The oscillation frequency and the displacement vector are written as the sum of unperturbed and perturbed quantities, i.e. $\omega= \omega_0 + \omega_1, \vxi = \vxi_0 + \xi_1$, with 
$\omega_1 \ll \omega_0$ and $\|\vxi_1 \ll \vxi_0\|$, and inserted into Eq.~(\ref{start}). 
At zero order, the system reduces to the eigenvalue problem governing the oscillations of a rotating non-magnetic star :
\begin{equation}
\omega_0^2 \vec{\xi}_0 - 2 i \omega_0 \vec{\Omega} \wedge \vec{\xi}_0 = {\cal L}_0(\vec{\xi}_0).
\end{equation}
In the low frequency domain, this problem can simplified using the TAR approximation \citep{E60,LS97,T03}. In this framework,
the centrifugal force, the perturbations of the gravitational potential and the latitudinal component of the
rotation vector $\vec{\Omega}$ are neglected. The eigenvalue problem then becomes separable in the spherical coordinates so that
the eigenfunctions $\vec{\xi}_0$ can be written as :
\begin{equation}
 \vec{\xi}_0= e^{i(m \phi + \omega_0 t)} \left(\xi_r(r) H_r(\theta) \er + \xi_h(r) H_\theta(\theta) \etheta + i \xi_h(r) H_\phi(\theta) \ephi \right) 
\end{equation}
where the variations in latitude are given by the three functions, $H_r, H_\theta, H_\phi$, which are known as Hough functions. The radial Hough function, $H_r$, is 
the solution of the Laplace's tidal equation :
\begin{equation}
\label{Laplace}
{\cal L_{\rm T}} (H_r) + \Lambda H_r =0 
\end{equation}
\noindent with
\begin{align*}
\label{Laplace_b}
{\cal L_{\rm T}}(H_r) =& \frac{d}{d \mu} \left[\frac{1-\mu^2}{(1-s^2\mu^2)} \frac{d}{d \mu} H_r \right] \\
& - \frac{1}{1-s^2\mu^2} \left( \frac{m^2}{1-\mu^2} + ms\frac{1+s^2\mu^2}{1-s^2\mu^2} \right) H_r 
\end{align*}
where $\mu = \cos \theta$, $\theta$ is the colatitude, and $s=\frac{2 \Omega}{\omega_0}$ is the spin parameter.
For a non-rotating star, the eigenvalues $\Lambda$ are equal to $\ell(\ell+1)$ for all azimuthal numbers
$-\ell \leq m \leq \ell$ and the eigenfunctions are a linear combination of the $2 \ell +1$ spherical harmonics $\YL$, $\ell$ denoting the degree of the spherical harmonic. The $m$-degeneracy is lifted in a rotating
star. The eigenvalues $\Lambda_{\ell,m}(s)$ and eigenfunctions $H_r^{\ell,m}(\theta,s)$ of the Laplace's tidal equation that correspond to g modes at vanishing rotation are labelled with $\ell$ and $m$. For the r modes that only exist in rotating stars a different label, $k$, is used instead of $\ell$. In the following the $\ell,m,s$ or $k,m,s$ dependence will be most often implicit to ease the reading. The horizontal Hough functions $H_\theta, H_\phi$ can be derived from $H_r$ through Eqs.~\ref{ht}, \ref{hp} given in Appendix \ref{aa}.

In the radial direction, the eigenvalue problem is identical to the radial eigenvalue problem of the non-rotating case except that $\ell(\ell+1)$ is replaced by $\Lambda$. Since $\gamma$ Dor stars oscillate in high-order  modes, the WKB solution of the radial eigenvalue problem provides a good approximate solution as long as the mode radial wavelength is much shorter than the scales of variation of the star structure. It reads : 
\begin{equation}
\label{asymp_g}
\frac{2 \pi}{\omega_0} = \frac{\Pi_0 (n + \epsilon_g)}{\sqrt{\Lambda}}
\end{equation}
with
\begin{equation}
\Pi_0 = \frac{2 \pi^2}{\int_{\ri}^{\ro} \frac{N}{r} \,\mathrm{d}r }
\end{equation}
\noindent $\ri$ and $\ro$, the inner and outer boundaries of the g-mode cavity, $n$ the mode radial order and $N(r)$ the Brunt-Väisälä frequency. As $\Lambda$ depends on $s$, Eq.~(\ref{asymp_g}) is an implicit equation for $s$. For a given $(\ell,m)$, it relates $s$ with $n$. 
In the following we shall also use the WKB solution of  horizontal component of the displacement vector : $\xi_h \propto \rho^{-1/2} r^{-3/2} N^{1/2} \Lambda^{-1/4} \sin(\Phi(r)) \omega^{-3/2}$
with $\Phi(r) = \int_{r_i}^r k_r -\pi/4 \; dr$ where $k_r^2 = \frac{N^2}{\omega_0^2} \frac{\Lambda}{r^2}$ represents the square of the local radial wavelength.

At first order, the eigenvalue problem Eq.~(\ref{start}) corresponds to the following eigenvalue problem for the perturbed frequency $\omega_1$ and the perturbed eigenfunctions $\vxi_1$ :
\begin{equation}
\omega_1 \left[ 2 \omega_0 \vxi_0 - 2 i \vec{\Omega} \wedge \vxi_0 \right] = \LopL(\vxi_0) - \omega_0^2 \vxi_1 + 2 i \omega_0 \vec{\Omega} \wedge \vxi_1 + \Lop_0(\vxi_1)
\end{equation}
As discussed in \citet{GT90} and \citet{MB23}, the modification of the hydrostatic equilibrium by the magnetic field is neglected as its effect on low frequency gravity modes is negligible compared to the effect of the perturbed Lorentz force $\LopL(\vxi_0)$.
Taking the scalar product of this equation with $\vxi_0$ and using the self-adjoint character of the operator $\Lop_0 + 2 i \omega \vec{\Omega} \wedge $, we
obtain :  

\begin{equation}
\label{adjoint}
\omega_1 = \frac{< \vxi_0, \LopL(\vxi_0) >}{2 \omega_0 < \vxi_0, \vxi_0> - < \vxi_0, 2 i \vec{\Omega} \wedge \vxi_0>}
\end{equation}

\noindent where the inner product is :
\begin{equation}
    \left\langle \vb{f}, \vb g \right\rangle = \int_V  \rho \vb f^* \cdot \vb g \ \;dV,
\end{equation}
with $V$ the star volume. This general expression of the magnetic shift was already derived in \citet{PM19}. 

In the case of slowly rotating stars, \citet{HZ05} showed that the expression of the magnetic shift can be significantly simplified for low frequency g modes provided the magnetic field varies over scales larger than the mode radial wavelength. As detailed in Appendix~\ref{approx}, this simplification also holds for the low frequency gravito-inertial modes as long as $B_\phi/B_r \ll (N/\omega_0)$ and 
$B_\theta/B_r \ll (N/2\Omega)^{1/2} (N/\omega_0)^{1/2}$.  
In this regime, the magnetic frequency shift only involves the horizontal displacements and the radial field component :
\begin{equation}
\label{tar}
\omega_1 = \frac{1}{2 \mu_0\omega_0} \frac{\int_{r_i}^{r_o} \!\!\int_0^{\pi} | \frac{\partial}{\partial r}\!\! \lp r \xi_h \rp |^2 \lp H_\theta^2 + H_\phi^2 \rp \langle B_r^2 \rangle_\phi \sin \! \theta \; d\theta \; dr}{\int_{r_i}^{r_o} \! \rho r^2 \xi_h^2 \; dr \int_0^{\pi}\! \lp H_\theta^2 + H_\phi^2 - \frac{2 \Omega}{\omega_0} H_\theta H_\phi \cos \! \theta \rp \sin \! \theta \; d\theta}
\end{equation}
\noindent where
\begin{equation}
    \langle B_r^2 \rangle_\phi (r,\theta) = \frac{1}{2\pi} \int_0^{2\pi} B_r^2(r,\theta,\phi) d\phi. 
\end{equation}

This formula generalizes to an arbitrary magnetic field the expression obtained by \cite{PM19,PM20} for an oblique dipolar field. It is valid for low frequency (or equivalently high-order) TAR gravito-inertial modes in a magnetic field that is not strongly dominated by its horizontal components.

Using the WKB solution for $\xi_h(r)$ and the stationary phase approximation, the integral involving $\xi_h(r)$ can be simplified as :
\begin{equation}
\frac{\int_{r_i}^{r_o} | \frac{\partial}{\partial r} \lp r \xi_h(r) \rp |^2 f(r) \; dr}{2 \int_{r_i}^{r_o} \rho r^2 \xi_h^2 \; dr} = \frac{\Lambda}{2 \omega_0^2}  
\frac{\int_{r_i}^{r_o} \lp \frac{N}{r} \rp^3 \frac{f(r)}{\rho} \; dr}{\int_{r_i}^{r_o} \frac{N}{r} \; dr}
\end{equation}
where $f(r)$ is an arbitrary smooth function. The magnetic frequency shift then becomes:

\begin{align}
\label{one}
\omega_1 =& \frac{1}{\mu_0} \frac{\Lambda}{2 \omega_0^3} {\cal I} \frac{\int_0^\pi \lp H_\theta^2 + H_\phi^2 \rp \sin \theta \; d\theta}{\int_0^{\pi} \lp H_\theta^2 + H_\phi^2 - \frac{2 \Omega}{\omega_0} H_\theta H_\phi \cos \theta \rp \sin \theta \; d\theta} \nonumber \\
&\times \int_{r_i}^{r_o} \!\! \int_0^\pi K_r(r) K_\theta(\theta) \langle B_r^2 \rangle_\phi \sin \theta \; d\theta \; dr
\end{align}
\noindent where 
\begin{align}
{\cal I} & =\frac{\int_{r_i}^{r_o} \frac{1}{\rho} \lp \frac{N}{r} \rp^3 \; dr}{\int_{r_i}^{r_o} \frac{N}{r} \; dr}
\\
K_r(r) & =  \frac{ \frac{1}{\rho} \lp \frac{N}{r} \rp^3 }{\int_{r_i}^{r_o} \frac{1}{\rho} \lp \frac{N}{r} \rp^3 \; dr} \\
K_\theta(\theta) & = \frac{ H_\theta^2 + H_\phi^2 }{\int_0^\pi \lp H_\theta^2 + H_\phi^2 \rp \sin \theta \; d\theta}.
\end{align}

The term ${\cal I}$ depends on the stellar structure within the oscillating cavity, $\langle B_r^2 \rangle_\phi$ is the azimuthal average of $B_r^2$, while $K_r$ and $K_\theta$ are respectively radial and latitudinal weight functions as they verify $\int_{r_i}^{r_o} K_r(r) \; dr = \int_0^\pi K_\theta(\theta) \sin \theta \; d\theta =1$.

In a non-rotating star, this expression reduces to :
\begin{equation}
\label{zero_rot}
\omega_1(s=0) = \frac{\cal I}{2\mu_0} \frac{\ell(\ell+1)}{\omega_0^3} \int_{r_i}^{r_o} K_r(r) \int_0^\pi K_\theta(\theta) \langle B_r^2 \rangle_\phi \sin \theta \; d\theta \; dr 
\end{equation}
where 
\begin{align}
K_\theta(\theta) =& \frac{ \lp\DYTL\rp^2+\frac{m^2}{\sin^2 \theta}\lp\YTL\rp^2 }{\int_0^\pi \lc \lp\DYTL\rp^2+\frac{m^2}{\sin^2 \theta}\lp\YTL\rp^2 \rc  \sin \theta \; d\theta} \nonumber \\
=& 
 \frac{2 \pi}{\ell(\ell+1)} \lc \lp\DYTL\rp^2+\frac{m^2}{\sin^2 \theta}\lp\YTL\rp^2 \rc
\end{align}
where $\YTL (\theta)$ is the latitudinal part of the spherical harmonic of degree $\ell$ and azimuthal number $m$, defined as $\YL(\theta,\phi)= \YTL (\theta) e^{im\phi}$. As expected, we recover the diagonal terms of the magnetic matrix derived by \cite{LD22}. We note that the non-diagonal terms
of the magnetic matrix are not relevant for the rapidly rotating stars considered in the present study. Indeed, they arise in slowly rotating from the coupling of the nearly degenerate multiplet components.  

To analyze the properties of the magnetic frequency shift in a rotating star, we find convenient
to rewrite Eq.~(\ref{tar}) as :
\begin{equation}
\label{two}
\omega_1 = \frac{\cal I}{2 \mu_0} \frac{F(s)}{\omega_0^3} T_{\!\!B}(s) \langle B_r^2 \rangle 
\end{equation}
\noindent where we introduce two dimensionless factors :
\begin{equation}
\label{FF}
F(s) = \Lambda \frac{\int_0^\pi \lp H_\theta^2 + H_\phi^2 \rp \sin \theta \; d\theta}{\int_0^{\pi} \lp H_\theta^2 + H_\phi^2 - s
H_\theta H_\phi \cos \theta \rp \sin \theta \; d\theta},
\end{equation}
\noindent and
\begin{equation}
\label{TT}
T_{\!\!B}(s) = \frac{\int_0^\pi K_\theta(\theta) \langle B_r^2 \rangle_{\phi,r} \sin \theta \; d\theta}{\langle B_r^2 \rangle},
\end{equation}
\noindent and two different averages of  $B_r^2$ :
\begin{equation}
\langle B_r^2 \rangle_{\phi,r}(\theta) = \int_{r_i}^{r_o} K_r(r) \langle B_r^2 \rangle_\phi \; dr,
\end{equation}
the average of $B_r^2$ in the azimuthal and radial directions, the average in the radial direction being weighted by $K_r(r)$ and performed between the inner and outer radii of the mode cavity, and
\begin{equation}
\langle B_r^2 \rangle = \frac{1}{4 \pi} \int_{r_i}^{r_o} K_r(r) \int_0^\pi \! \int_0^{2 \pi} B_r^2 \sin \theta \; d\theta \; d\phi \; dr
\end{equation}
the average of $B_r^2$ performed over the whole volume of the mode cavity with a $K_r$ weight in the radial direction. 

The factor $F$ depends on the mode, that is on $\ell$ (or $k$), $m$, and $s$, but not on the magnetic field or the stellar structure. 
The factor $T_{\!\!B}$ is a weighted average of $B_r^2$ scaled by $\langle B_r^2 \rangle$. It depends on the field topology but also on the mode through the latitudinal weight function $K_\theta(\theta)$. The product  ${\cal I} \langle B_r^2 \rangle$ depends on the stellar structure and on the mode considered through the radii of the mode cavity. 

\section{Analysis of the magnetic frequency shift for the modes identified in $\gamma$ Doradus stars}
\label{second}

In the following, we focus our analysis on the four types of modes that are most frequently identified
in $\gamma$ Doradus stars :
the two sectoral prograde g modes (also named Kelvin modes), $(\ell=1,m=-1)$ and $(\ell=2,m=-2)$, the r mode labelled $(k=-2,m=1)$ and the zonal dipolar
g mode $(\ell=1,m=0)$.
What is actually identified in these stars are frequency patterns formed by different radial orders $n$ of these modes.
Among the 611 $\gamma$ Dors analyzed by \cite{LV20},
594 stars show a frequency pattern of $(\ell,m) = (1,-1)$ modes, 172 a pattern of
$(\ell,m) = (2,-2)$ modes, 110 a pattern
of $(k=-2,m=1)$ modes
while 29 stars show a pattern of $(\ell,m) = (1,0)$ modes although only 11 stars among them have
typical $\gamma$ Doradus fast rotation ($P_{\rm rot} \sim 1 \; {\rm day}$). Retrograde dipolar modes have been also observed but only on slowly rotating stars and will not be considered here. The radial orders are rather high and the patterns cover large interval of radial orders. Indeed, for the two prograde sectoral modes, the radial order distribution has a median value of $n\sim 50$ and a mean pattern length of $30$ radial orders. For the $(k=-2,m=1)$ r modes, the median is lower, $n\sim 36$ but the mean pattern length is also $30$. Similar radial orders translate into different spin parameter intervals for the different modes. Indeed, for the prograde dipolar mode $(\ell=1,m=-1)$, the spin parameter distribution peaks at $s= 5$ and goes up to $\sim 20$ while it peaks at $s= 2.5$ and goes up to $\sim 10$ for the quadrupolar sectoral modes $(\ell=2,m=-2)$. The interval of $(k=-2,m=1)$ r mode spin parameters range from $6$ to $\sim 35$ with a peak at $s\sim9$. And among the 11 fast rotating stars with zonal dipolar
g mode $(\ell=1,m=0)$, the spin
parameter ranges from $\sim1$ to $\sim 2.5$.

For a given frequency pattern, that is for a given star and a fixed $(\ell,m)$ or $(k,m)$, the magnetic frequency shift varies with the spin parameter $s$ (or equivalently the radial order $n$).  In Sect.~\ref{sub1}, we study these variations and derive and test approximate analytical formulas for the frequency shift of the selected modes.  
In Sect. \ref{sub3}, we compare the magnetic shifts of these different modes and study a simple seismic diagnostic of the presence of magnetic fields when both  $(\ell,m) = (1,-1)$ and $(\ell,m)=(2,-2)$ modes are present. 
In Sect. \ref{sub2}, we construct synthetic frequency patterns affected by magnetic fields and test simple methods to detect magnetic signatures.

\subsection{Variations of the magnetic frequency shift with the spin parameter}
\label{sub1}

We successively study the terms $F$, $T_{\!\!B}$, and ${\cal I} \langle B_r^2 \rangle_{\phi,r}$ involved in the expression Eq.~(\ref{two}) of the frequency shift. To do so, we use both numerical and analytical calculations, the later being valid in the $s \gg 1$ limit. This allows us to derive and test approximate analytical formulas of the magnetic frequency shifts.

\subsubsection{The $F$ factor}

We recall that for a given $(\ell,m)$ the $F$ factor defined by Eq.~(\ref{FF}) only depends on $s$. 
The  numerical determination of $F$ requires to compute $\Lambda(s)$ and $H_r(\theta,s)$ from the Laplace's tidal equation, to derive $H_\theta$ and $H_\phi$ from ($\Lambda(s)$, $H_r(\theta,s)$), and to compute the integrals involved in the $F$ factor. These calculations are detailed in Appendix \ref{aa}. Approximate analytical expressions of $F$ can also be obtained from analytic expressions of $\Lambda$ and the Hough functions valid in the high-$s$ limit \citep{T03}. These calculations are detailed in Appendix~\ref{present} and \ref{F}.

\begin{figure}
\centering
\includegraphics[width=\hsize]{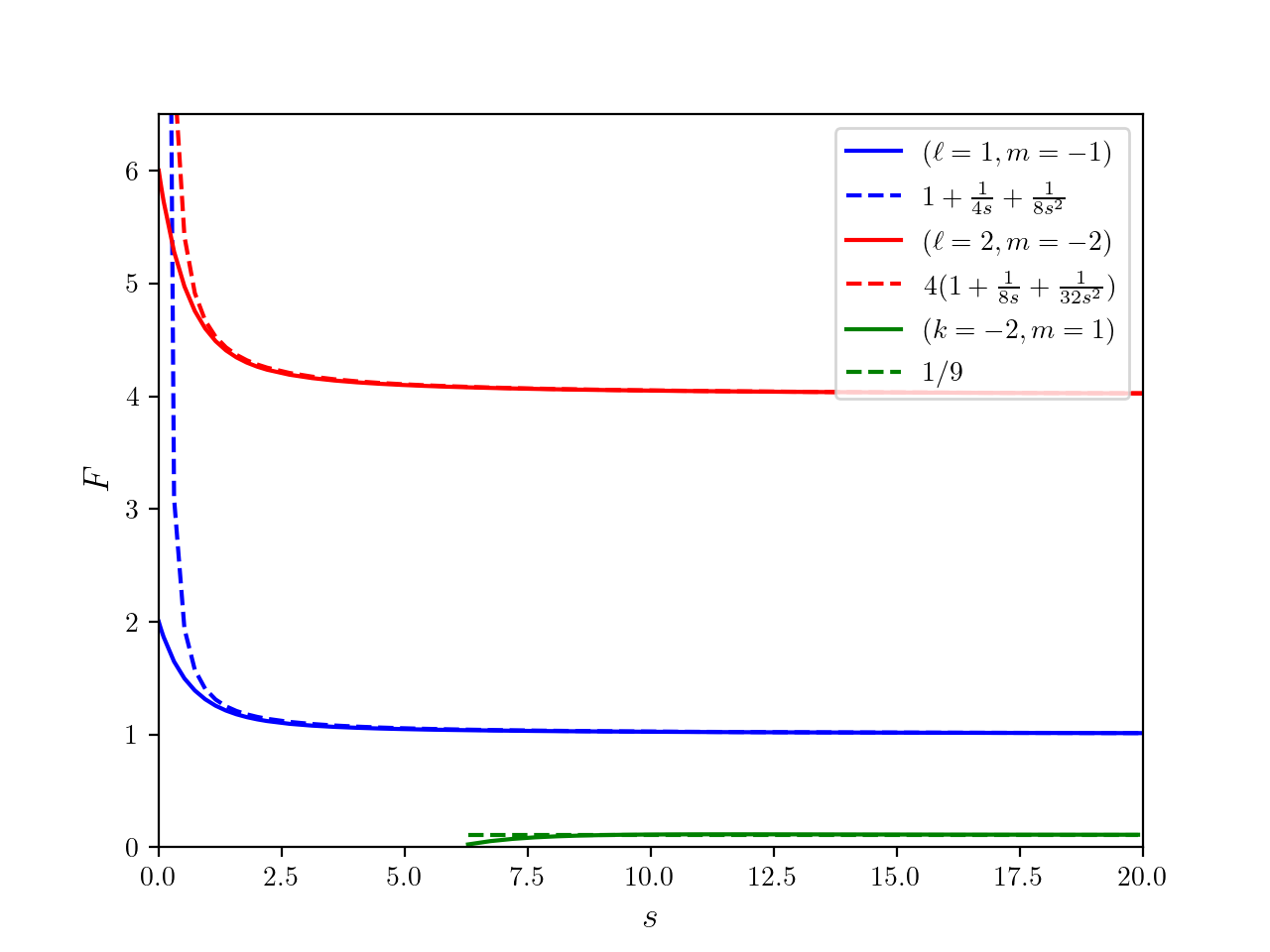}
\caption{The $F$ factor of the 
prograde sectoral modes $\ell = 1, m=-1$ (blue) and $\ell = 2, m=-2$ (red) and of the r mode  $k=-2, m=1$ (green) as a function of the spin parameter $s$. 
The asymptotic $s \gg 1$ forms
of the $F$ factor are  also plotted (dashed lines).}
         \label{sector}
\end{figure}

Figure~\ref{sector} displays both the numerical and the asymptotic forms of the factor $F$ for the two prograde sectoral modes $(\ell,m) = (1,-1)$ and $(\ell,m) = (2,-2)$ and for the r mode  $(k,m) = (-2,1)$.
As expected, $F(0) = \ell(\ell+1)$ for the two g modes when the rotation vanishes. As $s$ increases from $0$ to $20$, $F$ decreases by a factor $2$ for $(\ell,m) = (1,-1)$ and by a factor $2/3$ for $(\ell,m) = (2,-2)$.
$F$ actually approaches a constant value for large spin parameters, $F \sim 1$ for $(\ell,m) = (1,-1)$ and 
$F\sim 4$ for $(\ell,m) = (2,-2)$. This is in agreement with the analytical study that shows that $\lim_{s \rightarrow \infty} F_{|m|,m} = \lim_{s \rightarrow \infty} \Lambda_{|m|,m} = m^2$ for these two modes. More precisely, the
asymptotic forms are $F_{1,-1}(s) = 1 + 1/(4s) + 1/(8s^2)$ and  $F_{2,-2}(s) = 4(1 + 1/(8s) + 1/(32s^2))$ and, as shown on Fig.~\ref{sector}, they provide very good approximations when say $s\gtrsim 2.5$.

The r mode $(k,m) = (-2,1)$  only exists for spin parameters above $s = 6$ \citep{LS97}. Its $F$ factor vanishes there as
$\Lambda_{-2,1}(s=6) =0$. As for the two previous modes, $F_{-2,1}$ approaches $\Lambda_{-2,1} \approx 1/9$ for large $s$. For this mode, the high-$s$ analytical form is simply  $F_{-2,1} \approx 1/9$.
For the zonal dipolar mode $(\ell,m) = (1,0)$, the numerical and analytical $F$ factors 
are displayed in Figure~\ref{other} of Appendix \ref{bb}. The $F$ factor is very well approximated by its asymptotic form $F_{1,0}(s) =  (3/2)s^2$.

\subsubsection{The $T_{\!\!B}$ factor}

\begin{figure}
\centering
\includegraphics[width=\hsize]{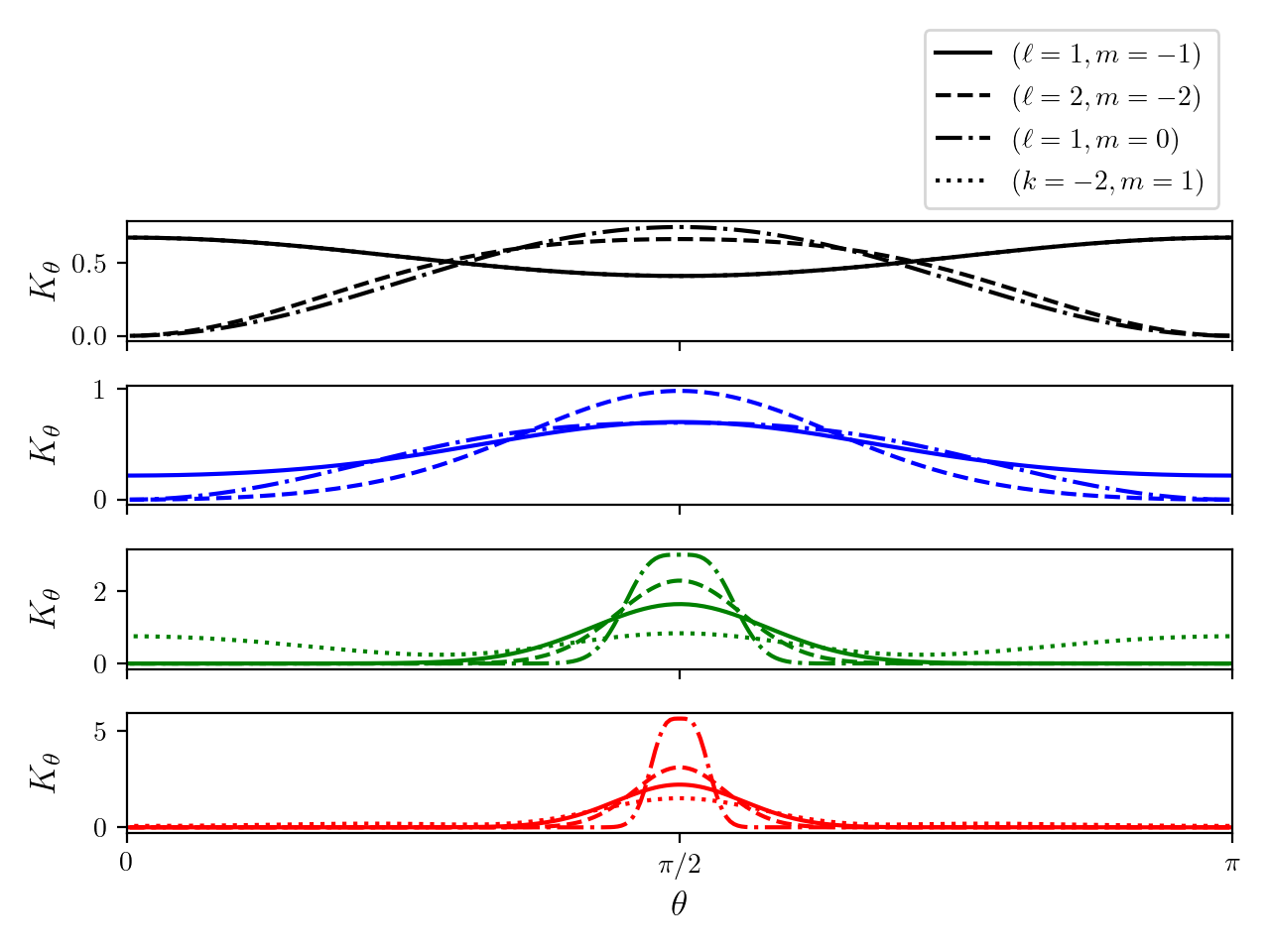}
\caption{The latitudinal weight functions $K_\theta$
of the two prograde sectoral modes : $\ell = 1, m=-1$ (continuous lines) and $\ell = 2, m=-2$ (dashed lines),
the zonal dipolar mode $\ell = 1, m=0$ (dashed-dotted lines) and the r mode $k= -2,  m=1$ (dotted lines) at four different spin parameters $s=0.1$ (black), $s =1$ (blue), $s=8$ (green), $s=15$ (red).}
         \label{allK}
\end{figure}

The factor $T_{\!\!B}$ defined by Eq.~(\ref{TT}) is a weighted latitudinal integral of
$\langle B_r^2 \rangle_{\phi,r}(\theta)$
normalized by  $\langle B_r^2 \rangle$. Here we 
analyze its variations with the spin parameter and the magnetic field topology for the same four modes.
For a given mode,  the weight function $K_\theta(\theta)$ only depends on the spin parameter. Figure~\ref{allK} shows the latitudinal profile of the weight function of the four modes computed at $s=[0.1,1,8,15]$.
We observe that  all the weight functions concentrate towards the equator as $s$ increases.
This is a consequence of the well-known tendency of the Hough functions to equatorial concentration \citep{LS97,T03}.
We also observe that at fixed $s$ this effect is more or less important depending on the mode. The axisymmetric dipolar mode is the most concentrated followed by the $(\ell,m) = (2,-2)$ mode, by  $(\ell,m) = (1,-1)$ mode and finally by the r mode which is not yet equatorially concentrated at $s=8$.  This behaviour is in qualitative agreement with the high-$s$ asymptotic form of the Hough functions \citep{T03} that indicate that the half-width of $H_\theta$ are
$1/s$ for the $(\ell,m) = (1,0)$ mode, $\sqrt{3/(-ms)}$ for the $(\ell,m) = (2,-2)$ and  $(\ell,m) = (1,-1)$ modes, and $\sqrt{3/(s-1)}$ for the r-mode.

As a consequence of the equatorial concentration, the $K_\theta$ average of $B_r^2$ 
is increasingly weighted towards the equator so that for very large spin parameters we expect  $T_{\!\!B} \approx
\langle B_r^2 \rangle_{\phi,r}(\theta = \pi/2)/\langle B_r^2 \rangle$. In this limit, $T_{\!\!B}$ is thus constant with $s$ and identical
for all the modes. From a Taylor expansion of $B_r^2$ at the equator, we can determine how $T_{\!\!B}(s)$ tends towards $\langle B_r^2 \rangle_{\phi,r}(\theta = \pi/2)/\langle B_r^2 \rangle$ (see details in Appendix~\ref{T}).
For the two sectoral prograde modes and the r mode, we obtain :
\begin{equation}
T_{\!\!B}^{\ell,m}(s) \langle B_r^2 \rangle = \langle B_r^2 \rangle_{\phi,r}(\theta=\pi/2) + \frac{1}{\alpha s} \frac{d^2 \langle B_r^2 \rangle_{\phi,r}}{d \theta^2}(\theta=\pi/2)
\end{equation}
\noindent with $\alpha =4$ for $(\ell=1,m=-1)$, $\alpha =8$ for $(\ell=2,m=-2)$ and $\alpha = 4/3$ for $(k=-2,m=1)$, while, for the dipolar mode axisymmetric,
\begin{equation}
T_{\!\!B}^{1,0}(s) \langle B_r^2 \rangle = \langle B_r^2 \rangle_{\phi,r}(\theta=\pi/2) + \frac{5}{12 s^2} \frac{d^2 \langle B_r^2 \rangle_{\phi,r}}{d \theta^2}(\theta=\pi/2).
\end{equation}
This approximation requires that near the equator $\langle B_r^2 \rangle_{\phi,r}(\theta)$ varies on scales larger than the width of the weight function $K_\theta$. To the first two orders in the $1/s$ expansion, $T_{\!\!B}$ depends on $\langle B_r^2 \rangle_{\phi,r}$ and its
second derivative at the equator. The next order would involve the fourth derivative of $\langle B_r^2 \rangle_{\phi,r}$ at the equator. These expressions also indicate that the convergence towards  $\langle B_r^2 \rangle_{\phi,r}(\theta = \pi/2)/\langle B_r^2 \rangle$ is faster ($\propto 1/s^2$) for the dipolar axisymmetric mode
than for the other three modes ($\propto 1/s$). 

\begin{figure*}
\centering
\includegraphics[width=\hsize]{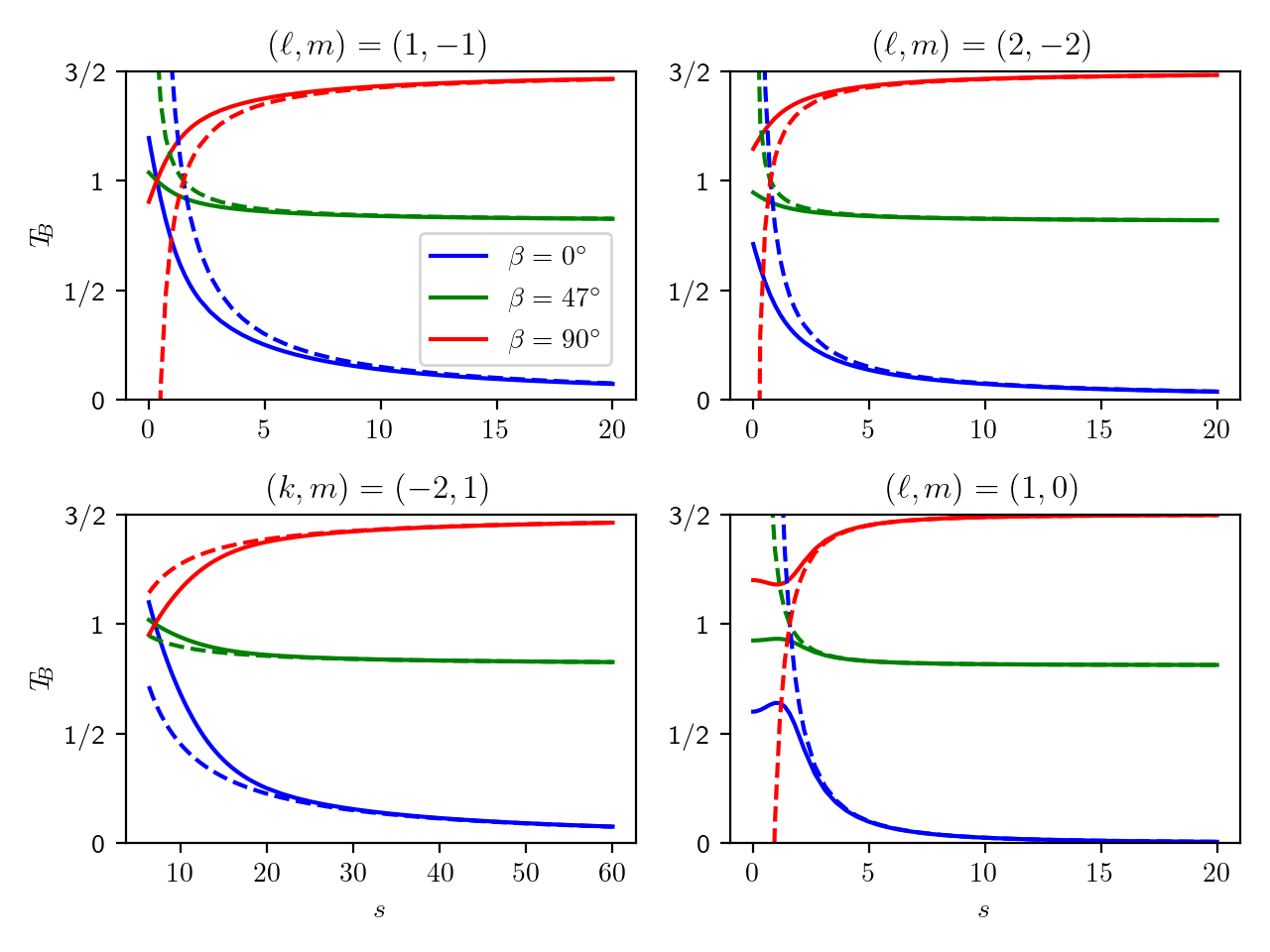}
\caption{The $T_{\!\!B}$ factor of the $(\ell, m) = [(1,-1),(2,-2),(1,0)], (k, m) = (-2,1)$ modes are displayed as a function of their spin parameter for three 
different angles of the inclined dipolar field $\beta =0^{\circ}$ (blue), $\beta=47^{\circ}$ (green),
$\beta=90^{\circ}$ (red). The continuous lines correspond to the numerical results and the dashed lines to the approximate analytical solutions valid in the high-$s$ limit.}
         \label{Tall_4}
\end{figure*}

To test these expressions, we  
computed $T_{\!\!B}(s)$ for an oblique dipolar field. This is a favorable case since the field varies on large scales around the equator. If $\beta$ denotes the inclination angle of the dipole with respect to the rotation axis, the radial component of the oblique dipole reads $B_r(r,\theta,\phi) = B_0 b(r) (\cos \theta \cos \beta + \sin \theta \cos \phi \sin \beta)$ \citep{PM20}. A straightforward calculation (see Appendix~\ref{od}) then yields :
\begin{align}
\label{Tod}
T_{\!\!B}(s) = & 3 \cos^2\beta\; \int_0^{\pi} K_\theta(\theta,s) \cos^2 \theta \sin\theta \; d\theta  \nonumber \\
+ &\frac{3}{2} \sin^2 \beta\; \int_0^{\pi} K_\theta(\theta,s) \sin^2 \theta \sin\theta \; d\theta
\end{align}
Figure~\ref{Tall_4}
shows both the numerical and analytical forms of $T_{\!\!B}(s)$ 
for the four modes and for three inclination angles $\beta= 0^\circ,47^\circ,90^\circ$ of the oblique dipole.  We find that the high-$s$ asymptotic expressions provide good approximations above some spin parameters, say above $s\gtrsim 5$ for $(\ell,m)=(1,-1)$, $s\gtrsim 2.5$ for $(\ell,m)=(2,-2)$ and  $(\ell,m)=(1,0)$ and $s\gtrsim 20$ for $(k,m)=(-2,1)$. As for the $F$ factor, the convergence is more or less rapid depending on the mode tendency to equatorial concentration. We also observe that, as expected,  $T_{\!\!B}(s)$ converge towards the same value $\langle B_r^2 \rangle_{\phi,r}(\theta = \pi/2)/\langle B_r^2 \rangle= \frac{3}{2} \sin^2 \beta$ for all the modes.

\subsubsection{The ${\cal I} \langle B_r^2 \rangle_{\phi,r}$  term}

\begin{figure}
\centering
\includegraphics[width=\hsize]{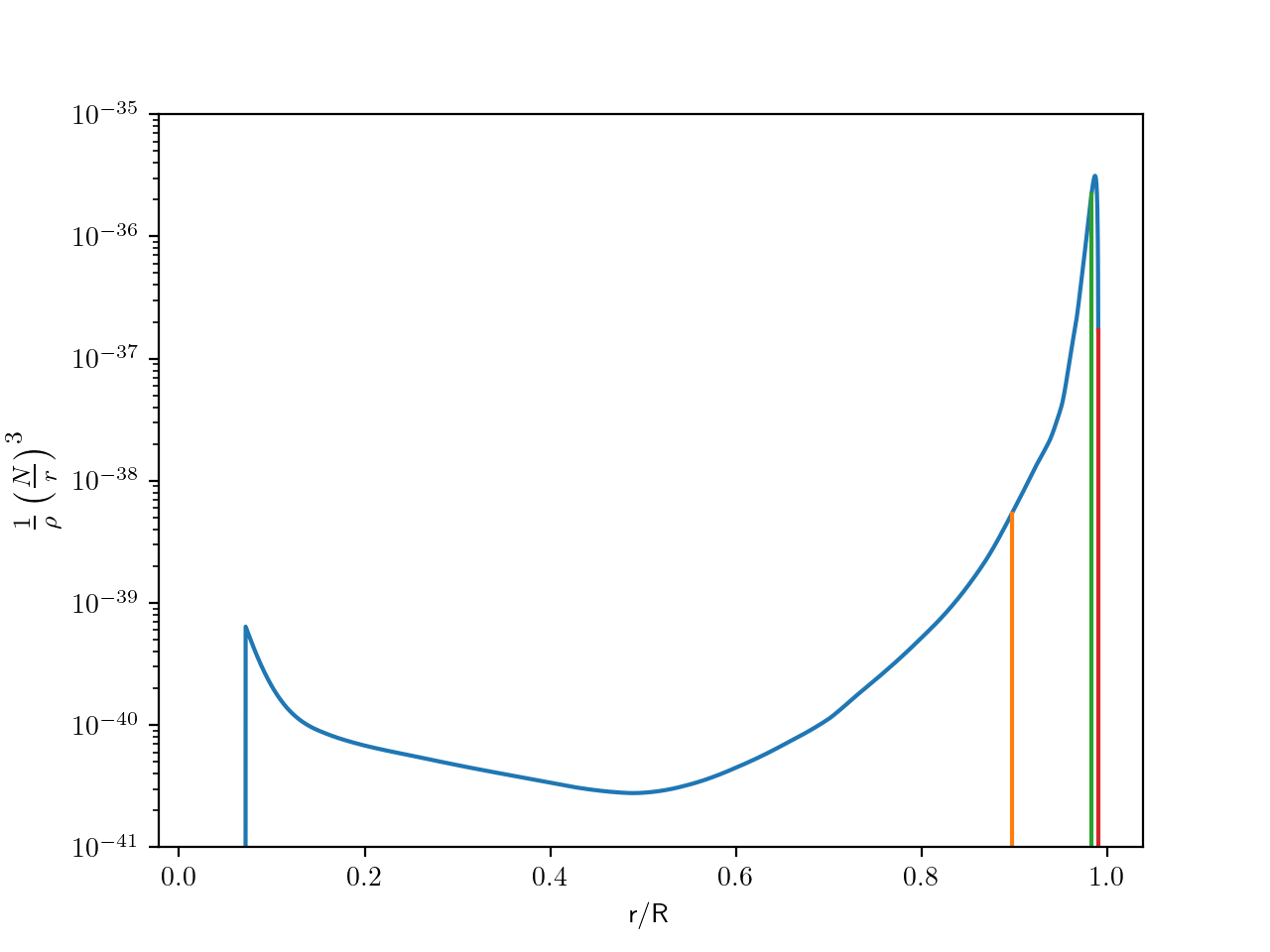}
\caption{The radial profile of $ \frac{1}{\rho} \lp \frac{N}{r} \rp^3 $ in c.g.s units for a M $=1.6$ M$_\sun$, R $=2.13$ R$_\sun$, X$_c = 0.35$ stellar model computed with the CESAM code. The outer radii $r_o$ of the oscillation cavity of $(\ell,m)=(1,-1)$ modes is shown by vertical lines for three different spin parameters $s=2$ (orange), $s=5$ (green) and $s=15$ (red).}
         \label{FKr}
\end{figure}

\begin{figure}
\centering
\includegraphics[width=\hsize]{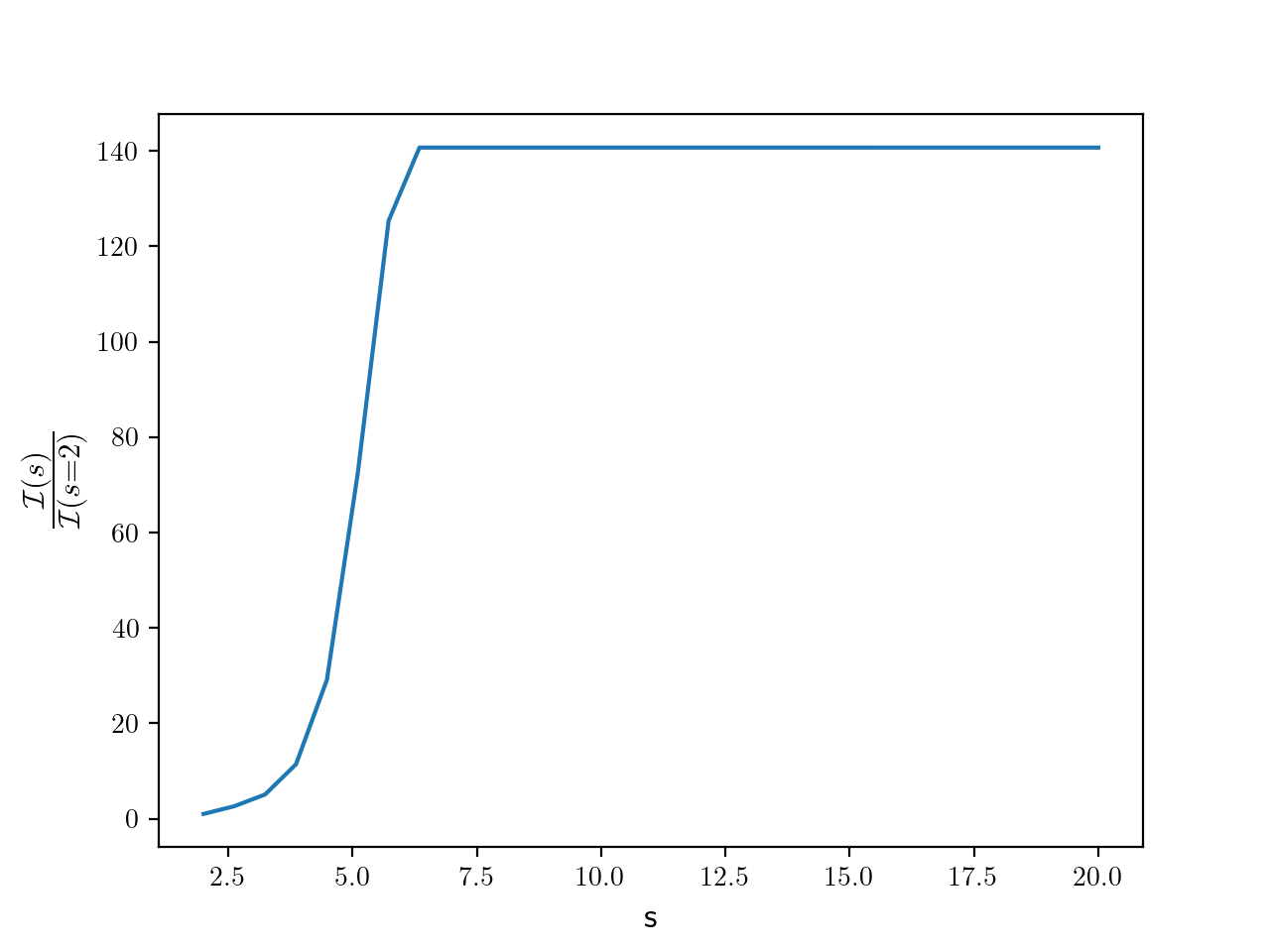}
\caption{Variation of ${\cal I} K_r(r)$ with $s$ computed for the $(\ell,m)=(1,-1)$ modes in the case of a uniform radial field (which implies that ${\cal I} K_r(r)  = {\cal I}$) and for a M $=1.6$ M$_\sun$, R $=2.13$ R$_\sun$, X$_c = 0.35$ stellar model.}
         \label{IKr}
\end{figure}

The magnetic frequency shift can also depend on $s$ through the radii of the mode cavity. This dependency is contained in the product of the two terms ${\cal I}$ and $ \langle B_r^2 \rangle_{\phi,r}$ : 
\begin{equation}
{\cal I} \langle B_r^2 \rangle_{\phi,r}  =
\frac{\int_{r_i}^{r_o} \frac{1}{\rho} \lp \frac{N}{r} \rp^3 \langle B_r^2 \rangle_\phi \; dr}{\int_{r_i}^{r_o} \frac{N}{r} \; dr}.
\end{equation}

The inner and outer cavity radii being determined by the conditions $\omega_0 = {\rm Min}(N,S)$, where $S = \frac{c_s}{r} \sqrt{\Lambda}$ is the Lamb frequency and $c_s$ the sound speed, they are likely to depend on the mode frequency $\omega_0$ and thus on $s$.
The inner radius $r_i$ remains very close to outer radius of the convective core for high-order g-modes though. This is due to the sharp increase of the Brunt-Väisälä frequency at the bottom of the radiative zone. Conversely, the outer radius $r_o$ which is located in the radiative envelope of $\gamma$ Dors does depend on the mode frequency. The vertical lines in Fig.~\ref{FKr} illustrate, in the case of the $(\ell,m)=(1,-1)$ mode, the variations of $r_o$ as $s$ takes three different values $[2, 5, 15]$

The variation of the outer radius of the mode cavity does not affect significantly $\int_{r_i}^{r_o} N/r \; dr$, the denominator of ${\cal I} \langle B_r^2 \rangle_{\phi,r}$, because $N/r$ is strongly peaked near the bottom of the radiative zone. 
But it can affect the numerator, that is the integral $\int_{r_i}^{r_o} \frac{1}{\rho} \lp \frac{N}{r} \rp^3 \langle B_r^2 \rangle_\phi \; dr$. Figure~\ref{FKr} indeed shows that $\frac{1}{\rho} \lp \frac{N}{r} \rp^3 $ is dominant and increases rapidly in the star envelope up to the surface convective zone where $N$ vanishes. Assuming a uniform magnetic field, Fig.~\ref{IKr} shows that this induces a sharp increase of ${\cal I} \langle B_r^2 \rangle_{\phi,r}$ 
between $s \sim 2$ and $s\sim 6$ for $(\ell,m)=(1,-1)$ modes.  For higher spin parameters, $r_o$ remains fixed at the bottom of the surface convective zone so that ${\cal I} \langle B_r^2 \rangle_{\phi,r}$ no longer varies with the spin parameter.
Considering a dipole-like $\propto 1/r^3$ decrease of $B_r$ within the envelope instead of a uniform field would not significantly modify this picture.
Conversely if the magnetic field decreases by various orders of magnitude between $r_i$ and $r_o$, the variations of $r_o$ will not affect the weighted integral of $\langle B_r^2 \rangle_{\phi}$. In this case, ${\cal I} \langle B_r^2 \rangle_{\phi,r}$  depends neither on $s$ nor on $(\ell,m)$.

\subsubsection{Approximate analytical formulas}

Using the analytical forms of $F$ and $T{\!\!B}$, we obtain the following approximate expressions for the magnetic frequency shifts of the four mode considered :
\begin{align}
\label{two_bis}
\omega_1^{1,-1} = &\frac{{\cal I} B_{eq}^2}{2 \mu_0} \frac{1}{\omega_0^3} \left( 1  + \frac{1 + D_{eq}}{4s} \right) \\
\label{two_bis_2}
\omega_1^{2,-2} = &\frac{{\cal I} B_{eq}^2}{2 \mu_0} \frac{4}{\omega_0^3} \left( 1 + \frac{1 + D_{eq}}{8s}  \right)  \\
\label{two_bis_3}
\omega_1^{-2,1} = &\frac{{\cal I} B_{eq}^2}{2 \mu_0} \frac{1}{9\omega_0^3} \left( 1 + \frac{3}{4} \frac{D_{eq}}{s}  \right)  \\
\label{two_bis_4}
\omega_1^{1,0} = &\frac{{\cal I}  B_{eq}^2}{2 \mu_0} \frac{6 \Omega^2}{\omega_0^5} \left(1 + \frac{5 D_{eq} }{12 s^2} \right)
\end{align}
\noindent where we introduced $B_{eq}^2 =  \langle B_r^2 \rangle_{\phi,r}(\theta = \pi/2)$ and $D_{eq} = \frac{d^2 \langle B_r^2 \rangle_{\phi,r}}{d \theta^2}(\theta=\pi/2)/B_{eq}^2$. These expressions are valid in the high-$s$ limit. They require that $\langle B_r^2 \rangle_{\phi,r}$ varies on a lengthscale larger than the half-width of the mode weight function $K_\theta$. They imply that in the high-$s$ limit these modes probe the magnetic fields only in the equatorial region. 

The analytical expressions allow us to conclude on the $s$ dependence of the frequency shift in the case of a magnetic field buried at the bottom of the radiative zone for which the product ${\cal I} B_{eq}^2$ has negligible variation with the spin parameter. 
Then, if $B_{eq}^2 \ne 0$, the magnetic frequency shift 
varies as $1/\omega_0^3$ for the two prograde sectoral modes and the r mode, and as $1/\omega_0^5$ for the 
dipolar axisymmetric modes. If $B_{eq}^2 =0$, the frequency shift rather varies as $1/\omega_0^2$ for the two prograde sectoral modes and the r mode, and as $1/\omega_0^3$ for the
dipolar axisymmetric modes. We should nevertheless keep in mind that the conditions $B_\phi/B_r \ll (N/\omega_0)$ and 
$B_\theta/B_r \ll (N/2\Omega)^{1/2} (N/\omega_0)^{1/2}$ are not verified at the equator in this case.

\begin{figure}
\centering
\includegraphics[width=\hsize]{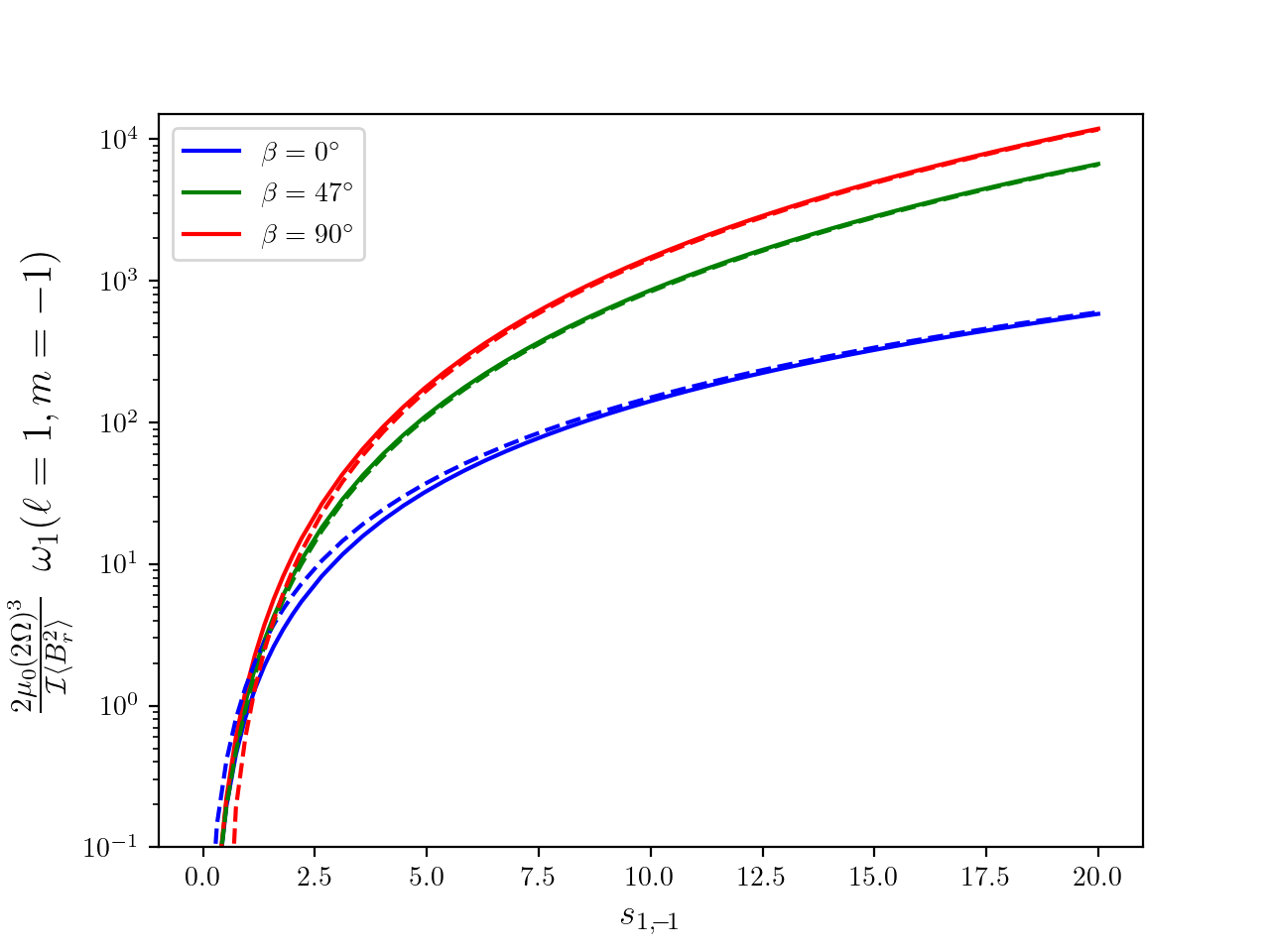}
\caption{The magnetic shift of $(\ell,m)=(1,-1)$ modes scaled by $\frac{{\cal I} \langle B_r^2 \rangle}{2 \mu_0 (2 \Omega)^3}$ calculated numerically (continuous lines) and analytically (dashes lines) for an oblique dipolar field . Three inclination angles of the dipole $\beta =0^{\circ}$ (blue), $\beta=47^{\circ}$ (green),
$\beta=90^{\circ}$ are considered.}
         \label{om1_s1}
\end{figure}

The magnetic shift scaled by $\frac{{\cal I} \langle B_r^2 \rangle}{2 \mu_0 (2 \Omega)^3}$, that is equal to $s^3 F(s) T_{\!\!B}$, is shown in  Fig.~\ref{om1_s1} for $(\ell,m)=(1,-1)$ modes in an oblique dipolar field. The numerical computation converges towards the analytical form for high spin parameters (the formula for oblique dipolar fields is derived in Appendix~\ref{od}). We see that the magnetic shift is larger for lower frequencies and strong equatorial radial fields. While the magnetic shift decreases at low $s$, it does not vanish in practice because $s$ has a lower limit associated with the upper limit of the g mode frequencies.  

We recall that if the envelope magnetic field is dominant, the variation of ${\cal I}\langle B_r^2 \rangle$ with $s$ (as shown in Fig.~\ref{IKr} for $(\ell,m)=(1,-1)$) must also be taken into account in the $s$ dependence of the frequency shift. 

\subsection{Comparing the magnetic shifts of different $(\ell, m)$ modes}
\label{sub3}

We now compare the magnetic shifts of different $(\ell, m)$ modes. This allows us to determine the mode that produces the highest shifts but also to find relations between the different magnetic shifts which 
can be used to detect/confirm magnetic signatures. 

Among the 611 $\gamma$ Dors studied by \citet{LV20}, 145 have both $(\ell=2, m=-2)$ and $(\ell=1, m=-1)$ frequency patterns identified, 
83 have both $(k=-2,m=1)$ r mode and  $(\ell=1, m=-1)$ mode frequency patterns identified and 27 have $(\ell=2, m=-2)$, $(\ell=1, m=-1)$,
and $(k=-2,m=1)$ frequency patterns identified. In addition,  
11 stars with typical $\gamma$ Dor rotation rates show both $(\ell=1, m=0)$ and $(\ell=1, m=-1)$ frequency patterns. The range of excited radial orders is similar for all these modes which prompts us to compare magnetic shifts of modes that have the same radial order $n$ but different horizontal quantum number $(\ell, m)$ or $(k,m)$.

We first compute and analyze the ratio of the magnetic shifts (Sect. ~\ref{sub3_1}) and then discuss how they can be used to detect magnetic signatures in frequency spectra (Sect. ~\ref{sub3_2}).

\subsubsection{The ratio of the magnetic shifts of different $(\ell, m)$ modes}
\label{sub3_1}

We consider modes having the same radial order $n$ but different horizontal quantum number $(\ell, m)$ or $(k,m)$. From the dispersion relation Eq.~(\ref{asymp_g}), it follows that the product $s \sqrt{\Lambda}$ is identical for all these modes :

\begin{equation}
\label{sl}
s \sqrt{\Lambda}= 2 \Omega \Pi_0 (n + \epsilon_g)/(2 \pi).
\end{equation}
Thus, knowing the spin parameter of say an $(\ell=1, m=-1)$ mode, $s_{1,-1}$, we can determine the spin parameters of the other modes by solving  the equation $ s \sqrt{\Lambda} = s_{1,-1} \sqrt{\Lambda_{1,-1}}$. This can be done numerically or using asymptotic forms of $\Lambda$ (see Eq.~(\ref{s_ratios}) in Appendix~\ref{cc}).
It is not necessary to specify the common radial order to determine the relations between the spin parameters, although $n$ can be determined
by Eq.~(\ref{sl}) once $\Omega$, $\Pi_0$ and $\epsilon_g$ are specified.

\begin{figure}
\centering
\includegraphics[width=\hsize]{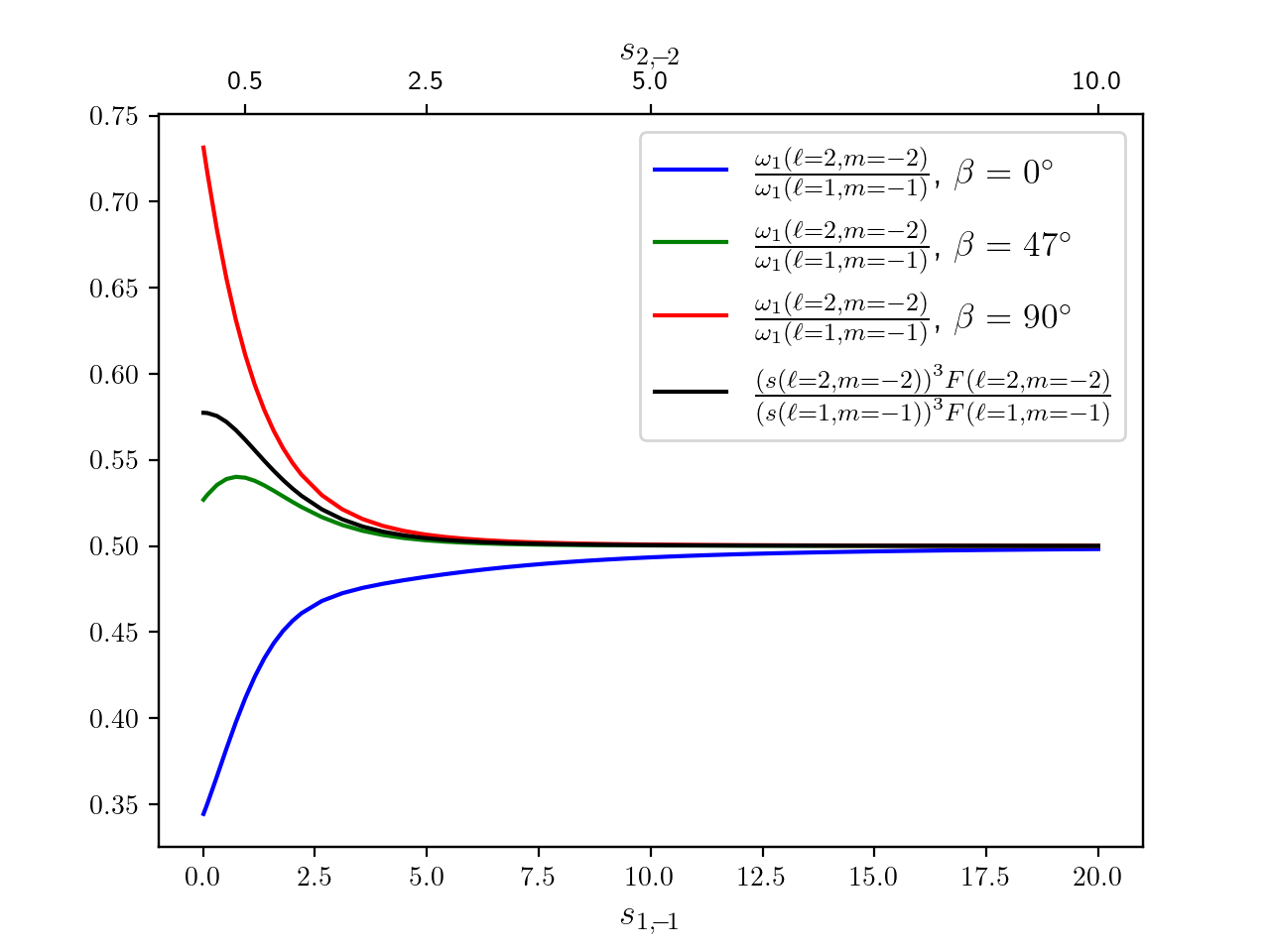}
\caption{Ratio of the  magnetic shifts produced by the prograde sectoral modes $\ell = 2, m=-2$ and $\ell = 1, m=-1$ of the same radial order $n$ as a function
of their spin parameters $s_{1,-1}$ (lower $x$ axis) and $s_{2,-2}$ (upper $x$ axis). The ratio is computed for three different angles of an inclined dipolar field, $\beta =0^{\circ}$ (blue), $\beta=47^{\circ}$ (green),
$\beta=90^{\circ}$. The ratio of the product $F(s)s^3$, which does not depend on the field topology, is also displayed (black).}
         \label{sectorR}
\end{figure}

For a given star, the ratio of the magnetic frequency shifts is equal to the ratio of $s^3F(s)T_{\!\!B}(s)$. The ${\cal I} \langle B_r^2 \rangle$ term cancels because the outer radii of the cavity is determined by $s \sqrt{\Lambda} = \frac{2 r\Omega}{c_s}$ and is thus identical for the different modes.  
The ratio between the magnetic shifts of the 
$(\ell=2, m=-2)$ and $(\ell=1, m=-1)$ modes, denoted $\omega_1(2, -2, n)/\omega_1(1, -1,n)$, is shown in Fig.~\ref{sectorR}.
It has been computed for oblique dipolar fields of three inclination angles and for the observed interval of prograde dipolar mode spin parameters, that is $s_{1,-1} \in [0,20]$, which 
translates into $s_{2,-2} \in [0,10]$ for constant $s \sqrt{\Lambda}$. The ratio of the $s^3F(s)$ terms, which does not depend on the field topology, is also displayed.

\begin{figure}
\centering
\includegraphics[width=\hsize]{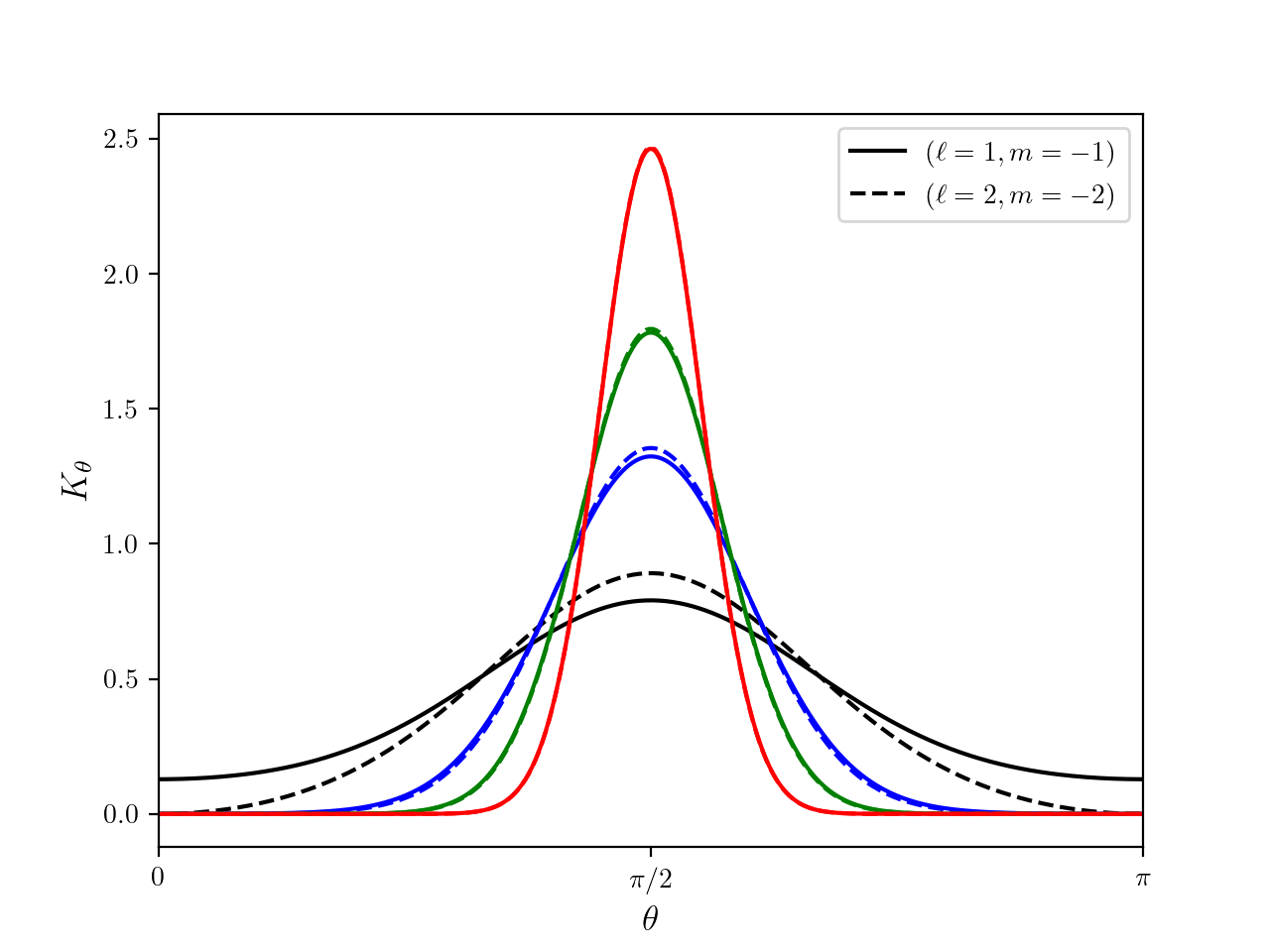}
\caption{Comparison of the latitudinal weight functions $K_\theta$ of
the $\ell = 1, m=-1$ (continuous lines) and $\ell = 2, m=-2$ (dashed lines) modes of same radial order $n$. The weight functions are computed for four $(s_{1,-1},s_{2,-2})$ couples : $(1.37,0.71)$ (black), $(4.94,2.47)$ (blue), $(9.5,4.75)$ (green), $(18.6,9.3)$ (red).}
         \label{Kth_compare}
\end{figure}

\begin{figure}
\centering
\includegraphics[width=\hsize]{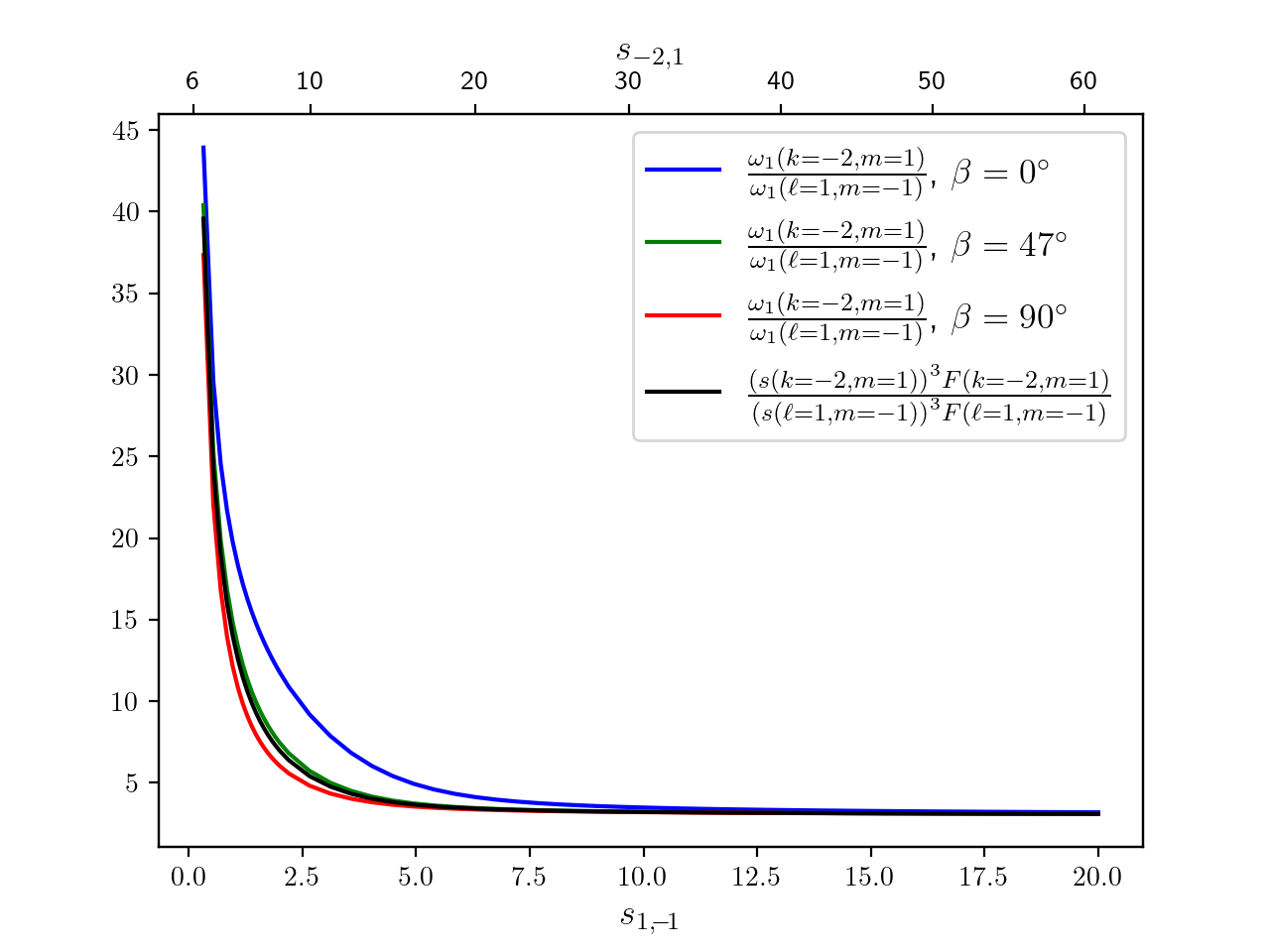}
\caption{Ratio of the  magnetic shifts of the r mode $k=-2, m=1$ and the prograde dipolar mode
$\ell = 1, m=-1$ of the same radial order $n$ as a function
of their spin parameters $s_{1,-1}$ (lower $x$ axis) and $s_{-2,1}$ (upper $x$ axis). The ratio is computed for three different angles of an inclined dipolar field, $\beta =0^{\circ}$ (blue), $\beta=47^{\circ}$ (green),
$\beta=90^{\circ}$. The ratio of the product $F(s)s^3$, which does not depend on the field topology, is also displayed (black).}
         \label{rossbyR}
\end{figure}

\begin{figure} 
\centering
\includegraphics[width=\hsize]{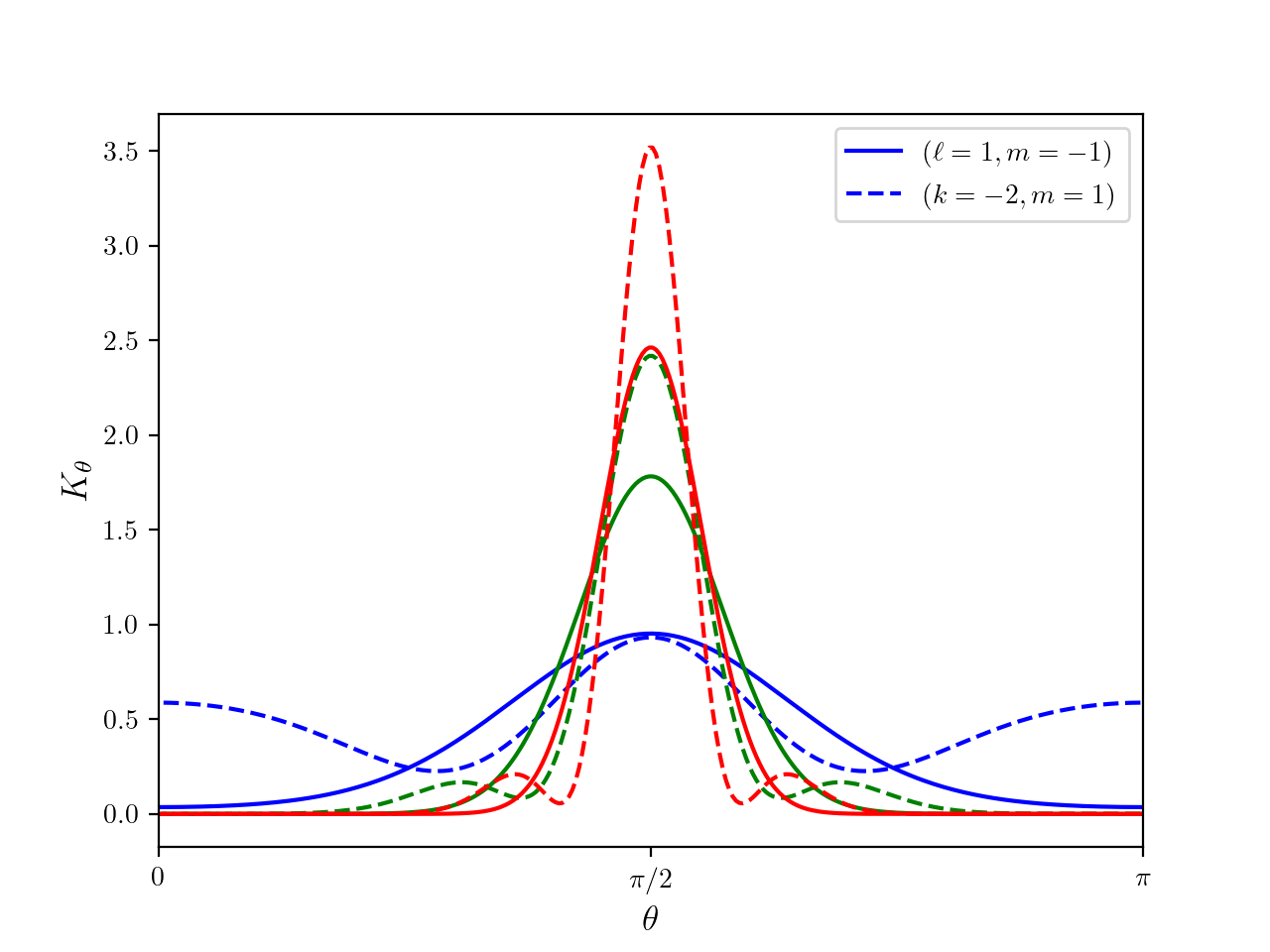}
\caption{Comparison of the latitudinal weight functions $K_\theta$ of
the $\ell = 1, m=-1$ (continuous lines) and $k=-2, m=1$ (dashed lines) modes of same radial order $n$. The weight functions are computed for three
$(s_{1,-1},s_{-2,1})$ couples : $(2.2,8.9)$ (blue), $(9.5,29.5)$ (green), $(18.6,56.8)$ (red).}
         \label{Kth_compare_bis}
\end{figure}

\begin{figure}
\centering
\includegraphics[width=\hsize]{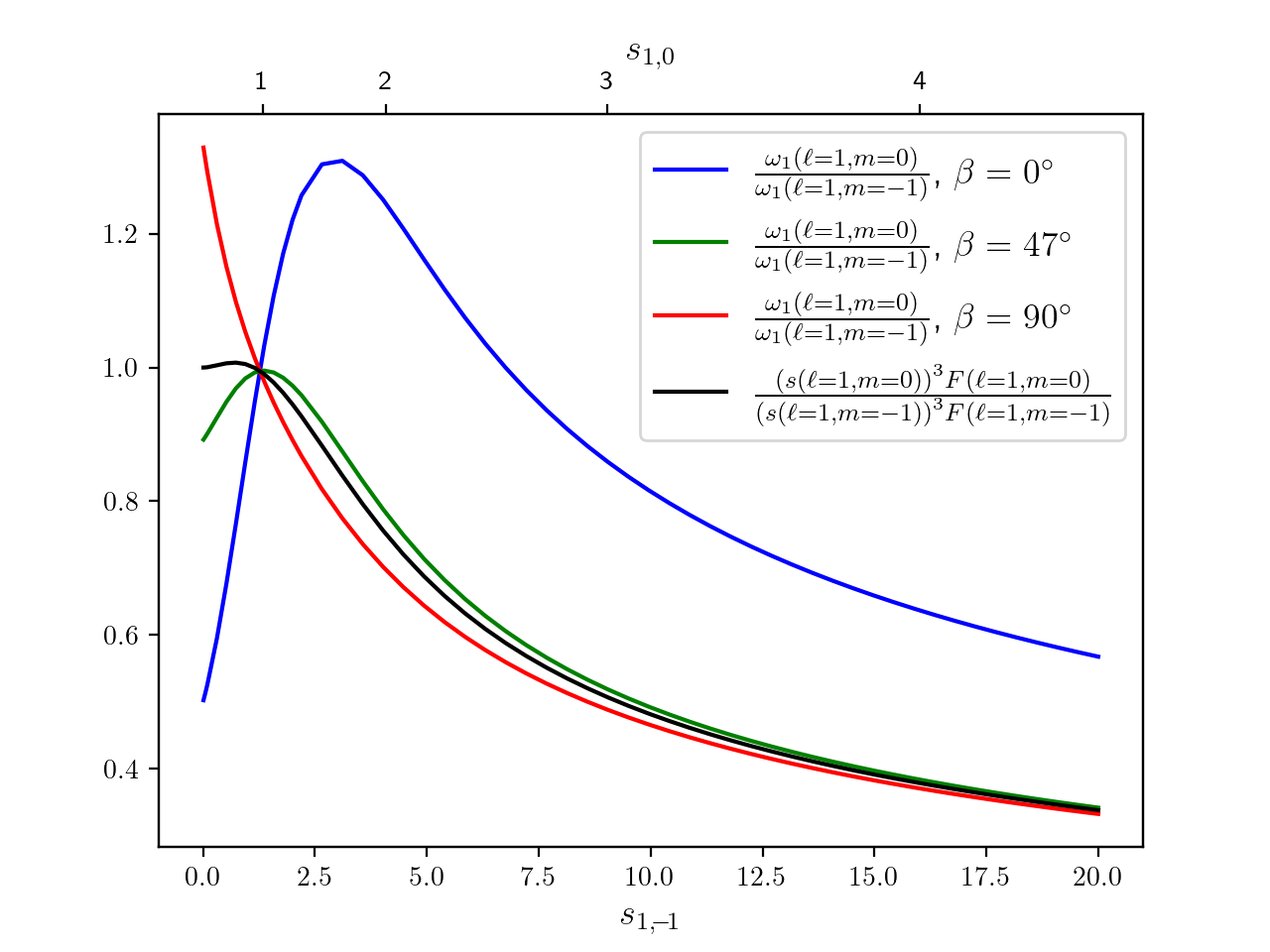}
\caption{Ratio of the magnetic shifts produced by the zonal dipolar mode $\ell = 1, m=0$ and the prograde dipolar mode $\ell = 1, m=-1$ of the same radial order $n$ as a function
of their spin parameters $s_{1,-1}$ (lower $x$ axis) and $s_{1,0}$ (upper $x$ axis). The ratio is computed for three different angles of an inclined dipolar field, $\beta =0^{\circ}$ (blue), $\beta=47^{\circ}$ (green),
$\beta=90^{\circ}$. The ratio of the product $F(s)s^3$, which does not depend on the field topology, is also displayed (black).}
         \label{axiR}
\end{figure}

\begin{figure}
\centering
\includegraphics[width=\hsize]{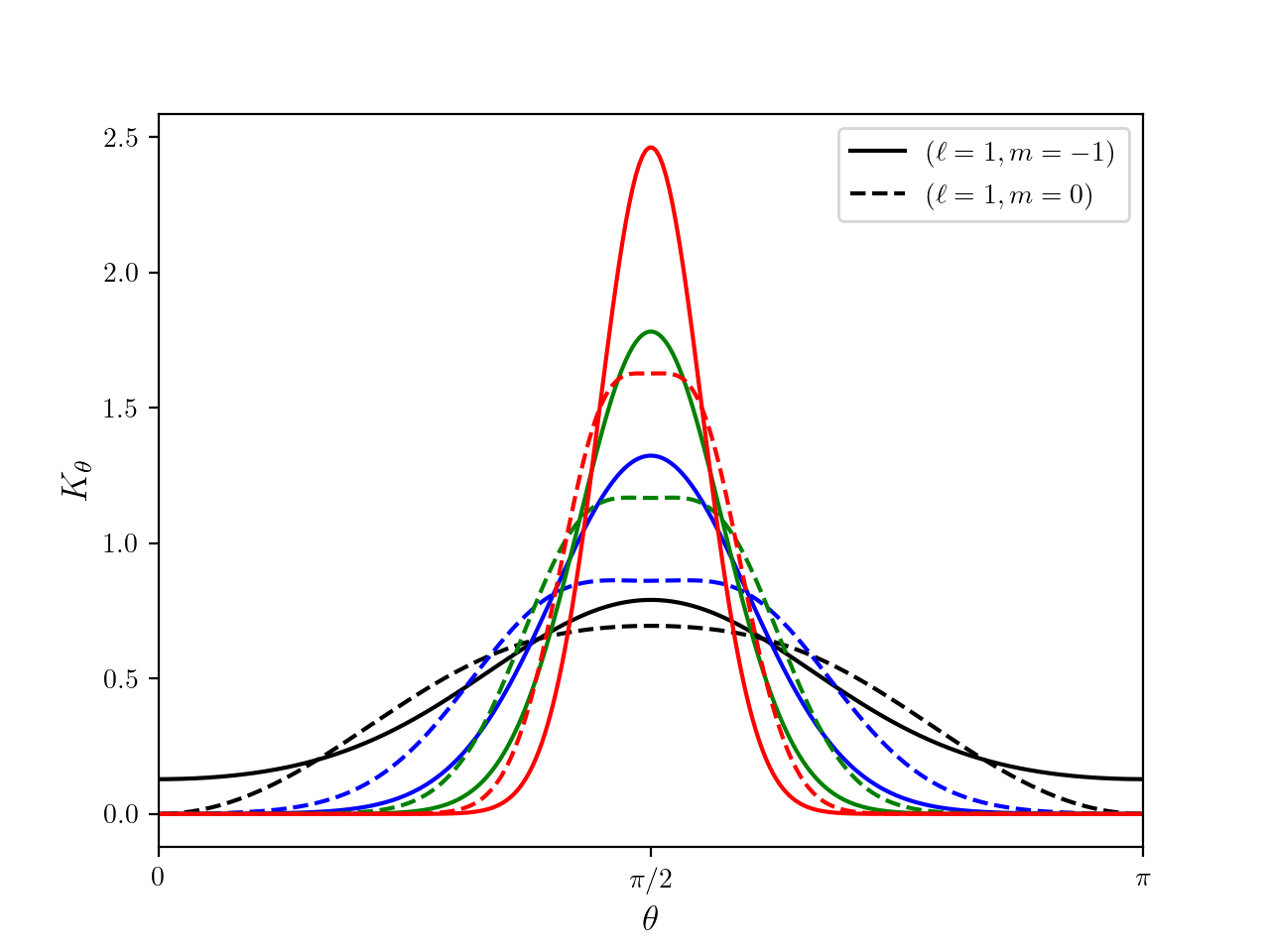}
\caption{Comparison of the latitudinal weight functions $K_\theta$ of
the $\ell = 1, m=-1$ (continuous lines) and $\ell=1, m=0$ (dashed lines) modes of same radial order $n$. The weight functions are computed for four
$(s_{1,-1},s_{1,0})$ couples : $(1.37,1.01)$ (black), $(4.94,2.21)$ (blue), $(9.5,3.08)$ (green), $(18.6,4.32)$ (red).}
         \label{Kth_compare_ter}
\end{figure}

We observe that the magnetic shift ratio $\omega_1(2, -2, n)/\omega_1(1,-1,n)$ is always smaller than $1$ 
and that, for all inclination angles, it converges rapidly towards $1/2$ as $s$ increases.
The asymptotic value of $1/2$ is predicted by the asymptotic form of the magnetic shift of the two modes (see Eqs.~(\ref{two_bis}) and~(\ref{two_bis_2})) and the asymptotic relation between the two spin parameters : $s_{2,-2}\approx s_{1,-1}/2$. The rapid convergence of $\omega_1(2, -2, n)/\omega_1(1,-1,n)$
can be understood by looking at the weight functions of the two prograde sectoral modes computed respectively at $s_{1,-1}$ and
$s_{2,-2}$. Figure \ref{Kth_compare} indeed shows that both functions become rapidly similar as $s$ increases above 1. This comes from the fact that at high $s$
the prograde sectoral horizontal Hough functions behave as Gaussian functions whose width is proportional to $1/(-ms)^{1/2}$.
Since $s_{2,-2}\approx s_{1,-1}/2$, 
the Gaussian widths of the two $(\ell = 2, m=-2, n)$ and $(\ell = 1, m=-1, n)$ Hough functions are very close, so that
we expect and verify that the weight functions $K_\theta^{1,-1}(s_{1,-1})$ and $K_\theta^{2,-2}(s_{2,-2})$ and therefore the $T_{\!\!B}$ factors of the two modes are also very close. 
Crucially, this implies that the property  $\omega_1(2,-2, n) \approx \omega_1(1, -1,n)/2$ holds for all field topologies but the rare ones that probe the small differences between the two latitudinal weight functions.

The ratio between the magnetic shifts of the r mode $(k=-2,m=1)$ mode and the prograde dipolar mode, $\omega_1(-2, 1, n)/\omega_1(1, -1,n)$ is shown in Fig.~\ref{rossbyR}.
The $s_{1,-1} \in [0,20]$ interval
translates into $s_{-2,1} \in [6,60]$  for constant $s \sqrt{\Lambda}$. As shown in Fig.~\ref{Kth_compare_bis}, the weight function $K_\theta$ of the two modes are strongly concentrated towards 
the equator at high $s$. The ratio of the magnetic frequency shifts tends towards $3$ in accordance with the analytical forms Eqs.~(\ref{two_bis}) and~(\ref{two_bis_3}) and the asymptotic relation $s_{-2,1}\approx 3 s_{-1,1}$.
We also observe that  $\omega_1(-2, 1, n)/\omega_1(1, -1,n)$ 
seems to diverge as $s_{-2,1}$ approaches $6$. Indeed,
at this point, $s_{1,-1}^3$ vanishes more rapidly than $F_{-2,2}$ so that the
ratio of $s^3 F(s)$ terms goes to infinity. 
This $s_{1,-1} \rightarrow 0$ limit is however not relevant in practice because both magnetic shifts vanish and will thus be undetectable.
It remains that, for $s_{1,-1} =2$, the ratio of the $s^3 F(s)$ terms is equal to $\sim 7$.
For these low  $s_{1,-1}$, the weight function profiles (see the blue curve in Fig.~\ref{Kth_compare_bis}) also shows that the r mode is more sensitive to fields concentrated in the pole 
than the dipolar sectoral mode. This gives an opportunity to probe magnetic fields in both polar and equatorial regions.

The ratio between the zonal and prograde dipolar mode magnetic shifts, $\omega_1(1, 0, n)/\omega_1(1, -1,n)$ is shown in Fig.~\ref{axiR}.
The $s_{1,-1} \in [0,20]$ interval                     
translates into $s_{1,0} \in [0,4.5]$  for constant $s \sqrt{\Lambda}$. The decrease of the ratio towards high $s$ is close to $3/(2s_{1,0})$ in agreement with the analytical forms Eqs.~(\ref{two_bis}) and~(\ref{two_bis_4}) and the asymptotic relation $s_{1,0} = \sqrt{s_{1,-1}}$. 
For $1 \lesssim s_{1,0} \lesssim 2.5$, the range of observed $ s_{1,0}$ in \citet{LV20},
the magnetic shift ratio is higher than $1$ if $\beta$ is low enough and is smaller than $1$ for higher $\beta$. 
Consequently, depending on the field topology, the rotational splitting 
$\omega_1(1, -1,n) - \omega_1(1, 0, n)$ can be either larger or smaller
than the unmagnetized TAR predictions. Such
rotational splittings have been determined by \citet{LV20}
for 11 stars and it was found that they can be either
higher or smaller than the TAR prediction. Magnetic fields are thus a potential candidate to explain these observations. 

\subsubsection{Using both $(\ell,m)=(1,-1)$ and $(\ell,m)=(2,-2)$ frequency patterns to detect and measure the magnetic frequency shift}
\label{sub3_2}

We now propose a method to search for and measure magnetic fields in stars where both $(\ell = 2, m=-2)$ and $(\ell = 1, m=-1)$ 
frequency patterns are present. If frequencies with the same radial order have been identified in these two frequency patterns, it is possible to compute the difference
between the $(\ell = 1, m=-1, n)$ frequency and half the  $(\ell = 2, m=-2, n)$ frequency, denoted $\delta K$ (as it involves two Kelvin modes). If magnetic effects can be treated perturbatively, it reads :
\begin{align}
\delta K_{\rm in} &= \nu_{\rm in}(1,-1,n) - \frac{\nu_{\rm in}(2,-2,n)}{2} \nonumber\\& = \nu_0(1,-1,n) - \frac{\nu_0(2, -2, n)}{2}  + \nu_1(1,-1,n) - \frac{\nu_1(2, -2, n)}{2} \\
& \approx \nu_0(1,-1,n) - \frac{\nu_0(2, -2, n)}{2}
+ \frac{3}{4} \nu_1(1, -1,n)  \label{latate1} \\
& \approx  \frac{3}{4} \nu_1(1,-1,n) \label{latate2}
\end{align}
The first approximated equation, Eq.~(\ref{latate1}), comes
from the relation $\omega_1(2,-2, n) \approx \omega_1(1, -1,n)/2$ found in Sec.~\ref{sub3_1}.  Eq.~(\ref{latate1}) then gives the magnetic shift $\nu_1(1, -1,n)$ if the same difference between the unperturbed frequencies
$\delta K_0 = \nu_0(1,-1,n) - \frac{\nu_0(2, -2, n)}{2}$ can be determined. This can be done in principle using a seismic determination of $\Omega$ and an unmagnetized oscillation model. But, we also expect $\delta K_0$ to be small and, if it is negligible compared to $\frac{3}{4} \nu_1(1, -1,n)$,  then Eq.~(\ref{latate2}) holds, giving a direct access to the magnetic frequency shift from the two observable frequencies $\nu_{\rm in}(1,-1,n)$ and $\nu_{\rm in}(2,-2,n)$.

\begin{figure}
\centering
\includegraphics[width=\hsize]{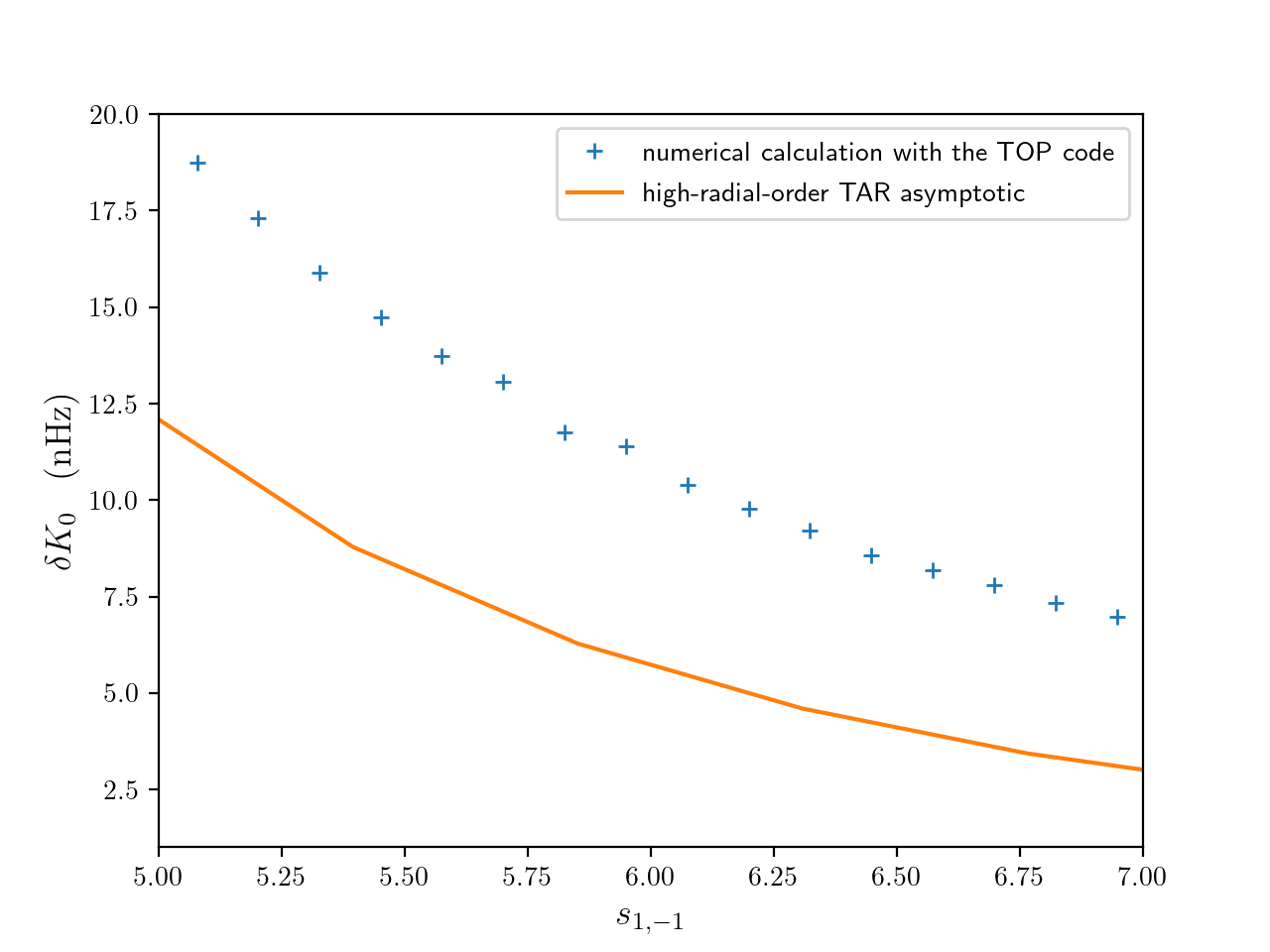}
\caption{The small difference $\delta K_0 = \nu_0(1,-1,n) - \nu_0(2, -2, n)/2$ computed in the $s_{1,-1} \in [5,7]$ interval for two models : a high-order TAR model at $\nu_{\rm rot} = 1.28$ d${}^{-1}$ (orange line) and $n \in [41,56]$ modes computed with the TOP code for a M $=1.4$ M$_\sun$, R $=1.38$ R$_\sun$, X$_c = 0.68$  stellar model rotating at $\nu_{\rm rot} = 1.28$ d${}^{-1}$ (blue crosses).}
         \label{halflaw}
\end{figure}

The difference $\delta K_0$ should indeed be small because 
it vanishes 
for high-$s$ high-order TAR modes (as $s_{2,-2} = s_{1,-1}/2$ in this limit). If we still assume high $n$ TAR modes but now consider finite $s$,  $\delta K_0$ is $ 9.6 $ nHz at $s_{1,-1}=5$ (for a rotation period of one day) and decreases for higher $s_{1,-1}$.
We also determined  $\delta K_0$ with more realistic frequencies computed from a stellar model and using the TOP code \citep{RL06,Reese09} to relax the TAR approximation of the Coriolis force. The structure of this model has been computed with the CESAM evolution code \citep{Morel97,Morel08}. It has a mass $ M =1.4$\,M$_\sun$, a radius $R=1.38$\,R$_\sun$, a central hydrogen abundance $X_c = 0.68$. The oscillation spectrum has been computed for a rotation $\nu_{\rm rot} =1.28$\,d$^{-1}$.
In Fig.~\ref{halflaw}, we show that $\delta K_0$ decreases from $\sim 20$\,nHz to $\sim 7$\,nHz between $s_{-1,1}=5$ and $s_{-1,1}=7$. This remains small although it is a bit higher than the same difference computed for high $n$ TAR modes.

While $\delta K_0$  tends to vanish as $n$ increases, the magnetic shift $\nu_1(1,-1,n)$ increases with $n$ (or $s$) as described by Eq.~(\ref{two_bis}). Thus, observing that $\delta K_{\rm in} \approx \delta K_0 + 3\nu_1(1,-1,n)/4$  
increases with $n$ is already a clue of the presence of a magnetic field. Fitting $\delta K_{\rm in}$ with Eq.~(\ref{two_bis}) then gives access to $B^2_{eq}$ and $D_{eq}$.

To test further the method proposed here, the properties of the difference $\delta K_0$ should be studied for various stellar models, including the effects of steep chemical gradients above the convective core, differential rotation and centrifugal deformation. These effects are briefly discussed in Sect.~\ref{disc}.  

\subsection{Probing magnetic signatures in frequency spectra}
\label{sub2}

In this section, we construct frequency spectra of prograde sectoral modes $({\ell =1, m=-1})$ and $({\ell =2, m=-2})$ affected by a radial magnetic field in order to test simple methods to reveal magnetic signatures. 

The spectra read $\nu_{\rm in} = \nu_{0} - m \nu_{\rm rot} + \nu_1$ where the 
index "${\rm in}$" indicates a frequency in the inertial frame, $\nu_0$ is the unperturbed frequency in the co-rotating frame and $\nu_1$ the magnetic frequency shift. The unperturbed frequency are computed using the TAR high-order approximation (Eq.~\ref{asymp_g}) with $\Lambda_{m,-|m|} = m^2\frac{2 m s}{2ms +1}$, $\nu_{\rm rot} = 1\,{\rm day}^{-1}$, and $\Pi_0 = 4122 \,{\rm s} $, for radial orders $n \in [35, 65]$.
The range of considered radial orders, the rotation frequency $\nu_{\rm rot}$ and the spacing period $\Pi_0$ correspond to typical values of the $\sim 700$ $(\ell,m) =(1,-1)$ and $(\ell,m) =(2,-2)$ patterns identified by \citet{LV20}.

This specific period $\Pi_0= 4122 \,{\rm s}$ has been calculated from a mid-main-sequence stellar model of a $\gamma$ Dor star ($M =1.6$ M$_\sun$, $R =2.13$ R$_\sun$, $X_c = 0.35$), which we use in this section. It has been computed with the CESAM evolution code, by including turbulent diffusion of chemicals $D=400$\,cm$^2$\,s$^{-1}$. It possesses a convective core with a radius $r_{cc}=0.069R$. 

The magnetic frequency shifts $\nu_1^{1,-1}$ and $\nu_1^{2,-2}$ are determined by Eqs.~\ref{two_bis} and ~\ref{two_bis_2} assuming $D_{eq} =0$ for simplicity. This stellar model and three different radial field profiles have been considered to compute ${\cal I} B^2_{eq}$ and then the magnetic frequency shifts.  
The first case (profile U) is a uniform field $B_r= 20$ kG between $r_{cc}$, the radius of the convective core, and $r=0.5 R$. 
For the two other profiles, we consider that $B_r$ worthes 24.3\,kG at $r=r_{cc}$ and decreases exponentially by one order of magnitude either at $r= 0.2 R$ (profile B+) or at $r= r_{cc} + 0.02 R$ (profile B-). 
The magnetic frequency shifts $\nu_1^{1,-1}$ computed for the frequency corresponding to $s_{1,-1}=5$ are $110$, $37$, $9.1$ ${\rm n Hz}$, for these three configurations respectively. The magnetic frequency shift $\nu_1^{2,-2}$ is simply $\nu_1^{1,-1}/2$ at $s_{2,-2}=s_{1,-1}/2$. 

These values are larger or just above the frequency resolution of \emph{Kepler} for classical pulsators ($1/T = 7.9$ nHz for $T=4$  years). 
In practice, models are not able to fit observed frequencies at this level of accuracy so that the magnetic frequency shifts should rather be compared with other source of frequency deviations such as the differential rotation or the steep gradient of molecular weight overlying above the convective core.

\begin{figure}[!ht]
\centering
\includegraphics[width=\hsize]{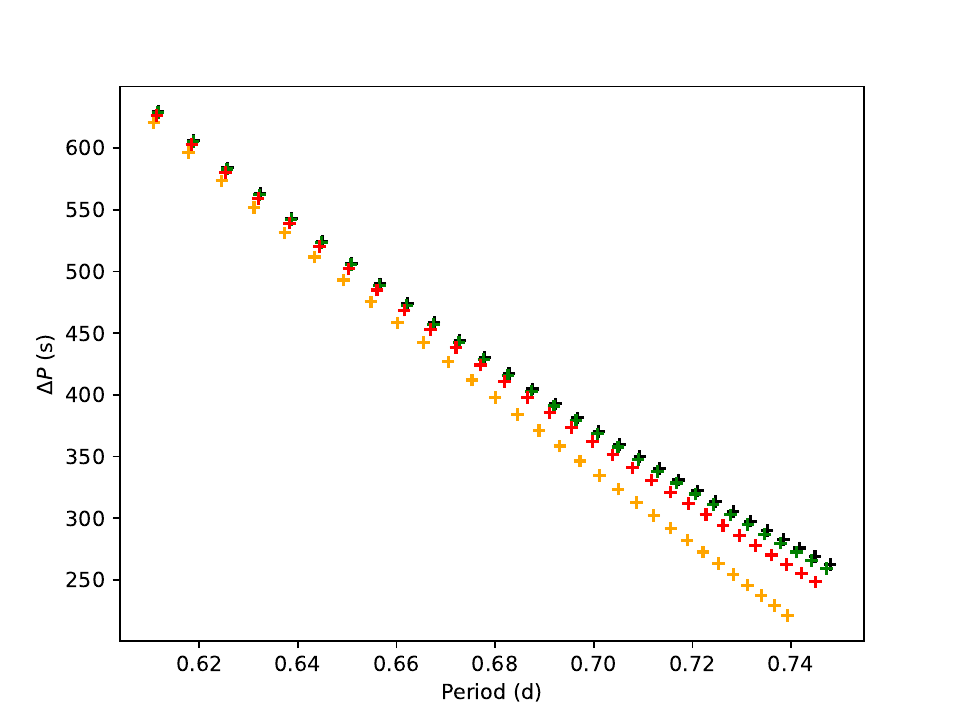}
\caption{Period spacing $\Delta P$ of $(\ell=1, m=-1)$ modes as a function of their periods for a model with a magnetic field following the radial profiles U (orange), B+ (red), and B- (green). The non-magnetic case is also shown (black).}
         \label{DP_P}
\end{figure}

\begin{table}[!ht]
    \caption{Values of $\Pi_0$, $\nu_\mathrm{rot}$, and $\epsilon_g$ fitted with the method of \citet{CB18} applied to frequency sets generated in presence of magnetic fields (profiles U, B+, B-). The table gives also the radial orders identified by the method. The first line corresponds to the real values without magnetic field.}
    \label{tab:fitres}
    \centering
    \begin{tabular}{lccccc}
    \hline\hline
          &               & $\Pi_0$ (s) & $\nu_\mathrm{rot}$ (d${}^{-1}$)  & $\epsilon_g$ & $n_\mathrm{min}$--$n_\mathrm{max}$\\
               \hline
         \multicolumn{2}{l}{real} 
                           & 4122 & 1.00 & 0.50 & 35--65\\
               \hline
          U    &  $\ell=1$ & 5309 & 1.07 & 0.26 & 30--60 \\
               &  $\ell=2$ & 4387 & 1.02 & 0.13 & 34--64 \\
          B+   &  $\ell=1$ & 4488 & 1.02 & 0.63 & 33--63 \\
               &  $\ell=2$ & 4208 & 1.01 & 0.04 & 35--65 \\
          B-   &  $\ell=1$ & 4209 & 1.01 & 0.04 & 35--65 \\
               &  $\ell=2$ & 4141 & 1.00 & 0.40 & 35--65 \\
               \hline
    \end{tabular}
\end{table}

\begin{figure}[!ht]
\centering
\includegraphics[width=\hsize]{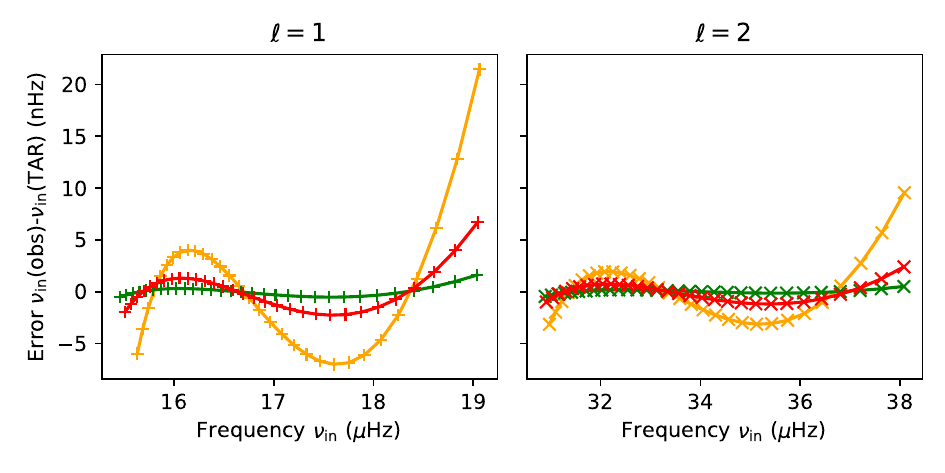}
\caption{Residuals between simulated frequencies and the fitted frequencies obtained with the method of \citet{CB18} applied to simulated data set with the profiles U (orange), B+ (red), and B- (green). The left (resp., right) panel shows results for $\ell=1$ (resp., $\ell=2$) modes.}
         \label{CB18_error}
\end{figure}
Figure~\ref{DP_P} shows the observed period spacings $\Delta P$ of $(\ell=1,m=-1)$ modes as a function of the period $P$ for the different magnetic cases. The $\Delta P - P$ diagrams are frequently used to determine the period $\Pi_0$ and the rotation frequency $\nu_\mathrm{rot}$ using the TAR \citep[e.g.][]{VT15,LV19}. Magnetic field does not produce noticeable signatures on the shape of the profile of $\Delta P$: it slightly changes its mean value and its slope. Such changes can be mimicked by a modification of $\Pi_0$ and/or $\nu_\mathrm{rot}$ and may bias the determination of these parameters.
To show this, we analyze the magnetically shifted frequency patterns with methods that do not take into magnetic effects. 
We thus apply the method developed by \citet{CB18} to the frequency sets of prograde dipolar and quadripolar modes generated with the magnetic field profiles U, B+ and B-. This method returns the period spacing $\Pi_0$ and the rotation frequency $\nu_\mathrm{rot}$ that fit the best a frequency set within the TAR. Once $\Pi_0$ and $\nu_\mathrm{rot}$ have been determined, we identify the radial orders $n$ of the modes and the offset $\epsilon_g$ (see Eq.~\ref{asymp_g}). Table~\ref{tab:fitres} summarizes the results obtained for the different cases and  Fig.~\ref{CB18_error} shows the difference between the simulated frequencies and the fitted frequencies.
First, we notice that the fits are in good agreement and the residuals between input and fitted frequencies do not generally exceed a few nHz. It means that a model without magnetic field can correctly reproduce data that are indeed affected by a magnetic field. However, the values of $\Pi_0$ and $\nu_{\rm rot}$ are biased towards larger values. The stronger the magnetic shift is, the stronger the bias is: $(\ell=1,m=-1)$ modes are more sensitive to magnetic field than $(\ell=2,m=-2)$ modes and the effect are the largest for profile U and the smallest for profile B-. The presence of a magnetic field also lowers the apparent value of the offset $\epsilon_g$. This effect can be strong enough to not only modify $\epsilon_g$, but also shift the apparent values of the radial order $n$ (it is the case for the profile U and for $\ell=1$ with the profile B+, see Table~\ref{tab:fitres}).

\begin{figure}[!ht]
\centering
\includegraphics[width=\hsize]{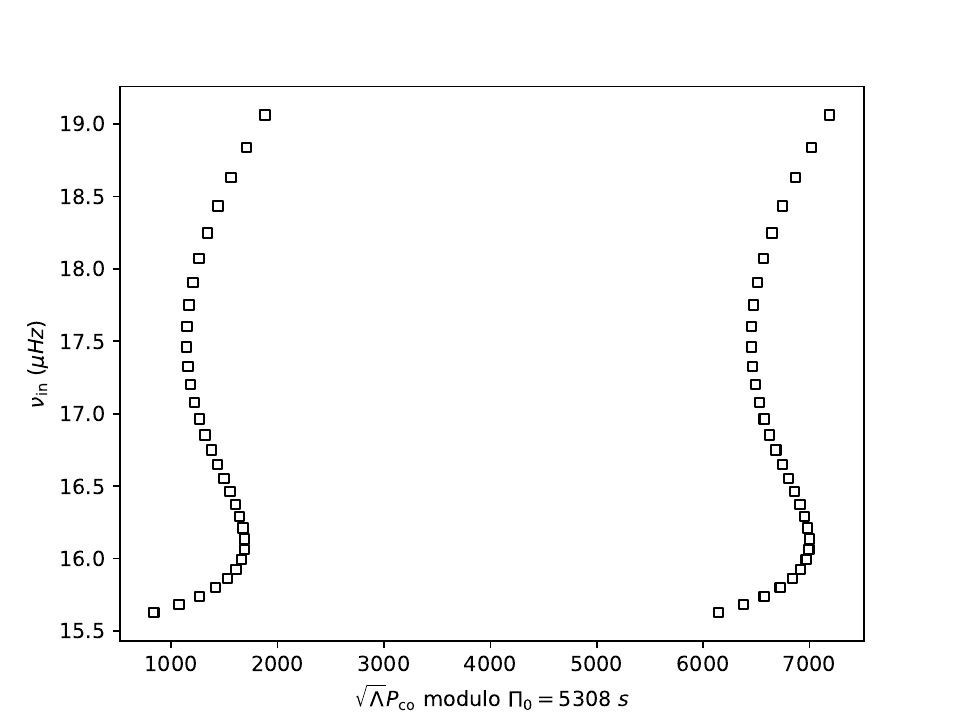}
\caption{Stretched échelle diagram from a magnetically shifted frequency spectra (profile U), obtained by applying the method of \citet{CB18}. Modes are plotted twice for clarity.}
         \label{CB18_SED}
\end{figure}
Considering a real star for which only $(\ell=1,m=-1)$ modes are observed, biases of $\Pi_0$ and $\nu_\mathrm{rot}$ can hardly be spotted since the real values of these quantities are a priori unknown.
However, for strong fields (such as with profile U, or with profile B+ to a lesser extent) we may identify the existence of a magnetic effect by analysing the departure of the observed frequencies from the TAR prediction. This difference can be revealed in a stretched échelle diagram, in which mode frequencies are plotted as a function of the period in the corotating frame, stretched with the $\Lambda$ function, modulo the period spacing $\Pi_0$ (Fig.~\ref{CB18_SED}). Within the TAR, modes form vertical lines in stretched échelle diagrams. We see in Fig.~\ref{CB18_SED} that a magnetic field distorts the ridge in a S-shape profile with an amplitude larger than 500\,s (for profile U). This signature can be compared to various effects studied in \citet{CB18}, especially in their Fig.~5. First we notice that the differences between complete computations and the TAR are clearly smaller (< 100\,s). Second, buoyancy glitches generate  oscillating signatures with amplitudes similar to the ones produced by a strong magnetic field, but their periodic nature helps to discriminate between both. Finally, a strong differential rotation also produces a smooth change in the ridge shape, but with a very small amplitude and mainly induces mode avoided crossings, which create a quite different signature compared to the one shown in Fig.~\ref{CB18_SED}.

Having both $(\ell=1,m=-1)$ and $(\ell=2,m=-2)$ frequency patterns helps detect seismic magnetic signatures. 
Table~\ref{tab:fitres} indeed shows clear differences in the values of $\Pi_0$ obtained for each pattern. For the U magnetic profile, the $(\ell=1,m=-1)$ pattern provides a period spacing 21\% larger than the $(\ell=2,m=-2)$ pattern. This corresponds to a difference of 922\,s that is larger than most errors in $\Pi_0$ determination quoted in \citet{LV20}. 
The difference in the $\nu_{\rm rot}$ values is smaller, although the 0.05\,d${}^{-1}$ discrepancy is also larger than most error in $\nu_{\rm rot}$ values quoted in \citet{LV20}. For the weakest magnetic shifts (profile B-), the difference in $\Pi_0$ is at the level of the best fitting errors indicated in \citet{LV20}.

\begin{figure}[!ht]
\centering
\includegraphics[width=\hsize]{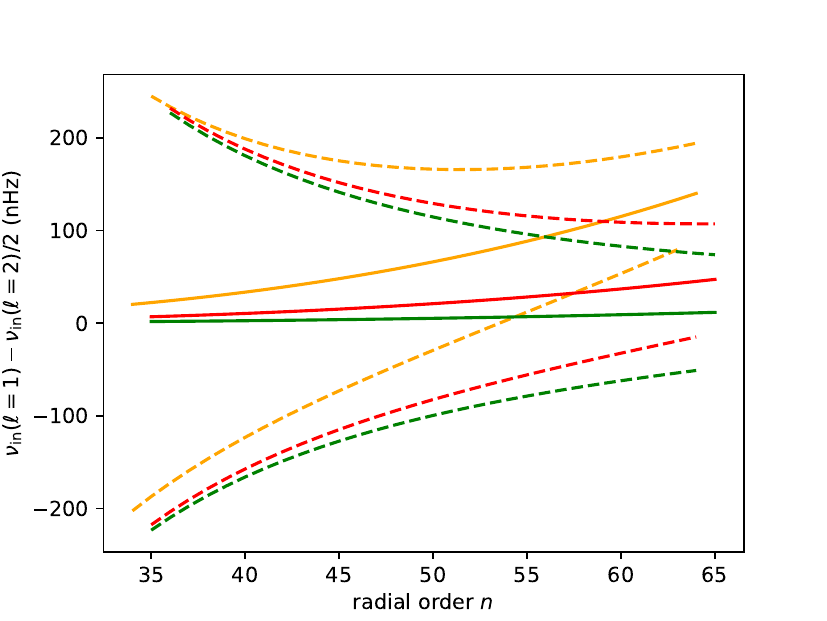}
\caption{Magnetic shift computed from the difference $\delta K_{\rm in}$ 
for the magnetic profile U (orange), B+ (red), and B- (green), as a function of the apparent radial order $n$ of $\ell=2$ modes. We notice that for the profile U, the apparent radial orders are shifted compared to the real ones (see table~\ref{tab:fitres} and text for a discussion). Dashed lines are obtained when the radial order of $\ell=1$ modes are incorrectly identified compared the the radial order of $\ell=2$ modes by $\pm 1$. }
         \label{shift_l1l2}
\end{figure}

The $(\ell=1,m=-1)$ and $(\ell=2,m=-2)$ frequency patterns can also be combined to detect and measure the magnetic frequency shifts. Using the radial order identifications provided by the fits (see Table~\ref{tab:fitres}), the difference $\delta K_{\rm in}$ introduced in Sec.~\ref{sub3_2} can be computed as a function $n$. In a non magnetic star $\delta K_{\rm in} = \delta K_0$ is small (of the order of 10\,nHz) and a decreasing function of $n$ whereas a magnetic term increasing with $n$ adds up in a magnetic star. As the order identifications may be wrong, differences formed by shifting the radial orders of one of the pattern $\nu_{\rm in} (\ell=1,m=1,n+n_s) - \nu_{\rm in} (\ell=2,m=2,n)/2$ are computed until the smallest positive difference is found. The
result of this process is shown in Fig.~\ref{shift_l1l2} for the three magnetic profile considered.  
We observe that the smallest positive difference is an increasing  function of $n$ for the three magnetic profiles. As explained in Sect.~\ref{sub3_2}, this quantity is close to $3\nu_1^{1,-1}/4$ and can be used to derive the magnetic seismic observables $B_{eq}$ and $D_{eq}$.  Here $\delta K_0 = 0$ because we computed the unperturbed frequencies in the asymptotic regime. In real stars, the magnetic term will need to be higher than $\delta K_0$ for the smallest positive difference $\nu_{\rm in} (\ell=1,m=1,n+n_s) - \nu_{\rm in} (\ell=2,m=2,n)/2$ to be an increasing function of $n$. Considering that $\delta K_0$ is of the order of 10\,nHz as in the realistic model considered in Sect.~\ref{sub3_2} (see Fig.~\ref{halflaw}), a magnetic signature would be detected for the magnetic profiles U and B+.

\section{Summary and discussion}
\label{disc}

We used a perturbative method to investigate the frequency shift induced by an arbitrary magnetic field on gravito-inertial modes. From the general expression of the magnetic frequency shift, Eq.~(\ref{adjoint}), we derived approximated simplified formulas valid for short radial wavelength gravito-inertial modes described by the TAR. This allows us to determine the frequency dependence of the magnetic shift for the modes most often observed in $\gamma$ Dor stars, the $(\ell,m) =(1,-1),(2,-2),(1,0)$ g modes and the $(k,m) =(-2,1)$ r mode, and to define the seismic observable quantities and their relation with the internal magnetic field.  
We find that the magnetic frequency shift is larger for low frequency
modes and for strong equatorial radial fields. Comparing magnetic 
shifts of different modes with the same radial order $n$, we find that the ($k=-2,m=1$) r mode is the most sensitive to magnetic field followed by the $(\ell =1,m=-1)$ g mode.  We then proposed simple methods to detect magnetic signatures : the characteristic form of the error when the frequency pattern is analyzed with a non-magnetic model, the discrepancy between the period spacings $\Pi_0$ determined respectively with the $(\ell,m)=(1,-1)$ and  the $(\ell,m)=(2,-2)$ frequency patterns analyzed with a non-magnetic model, and the determination of the magnetic shift from the difference $\delta K_{\rm in} = \nu_{\rm in}(1,-1,n) - \nu_{\rm in}(2,-2,n)/2$.

Before starting the discussion of our results, we first have to mention a clear discrepancy between our results and those presented in \citet{PM19,PM20}. The magnetic shifts computed in these papers result in a sawtooth pattern in the period spacing vs period relation. This feature is incompatible with the smooth variations of the magnetic shift with $s$ (or $n$) that we find in both our numerical and analytical results. We notice that in \citet{PM19} the equations relating the horizontal Hough functions to the radial Hough function given in their Appendix A are not correct.

In the following we first discuss the approximations made to derive the simplified formulas of the magnetic frequency shifts, then the conditions for applying the perturbative method, and finally the origin and detectability of magnetic fields in $\gamma$ Dor radiative zones.

\subsection{Approximations in the derivation of the magnetic shifts}

In the regime of short radial wavelengths, the effect of the radial component of the magnetic field tends to dominate over the horizontal components. As already discussed in \citet{LD22}, the contribution of the horizontal components to the magnetic shift can be neglected if 
$k_r B_r \gg k_\theta B_\theta$ and  
$k_r B_r \gg k_\phi B_\phi$, where 
$(k_r,k_\theta,k_\phi)$ are the components of the mode wavevector. When applied to high-$n$ high-$s$ TAR oscillations, these conditions translate into upper limits of the ratio of the horizontal to radial components namely 
$B_\phi/B_r \ll (N/\omega_0)$ and 
$B_\theta/B_r \ll (N/2\Omega)^{1/2} (N/\omega_0)^{1/2}$  
(see details in Appendix~\ref{approx}). 
For a $(\ell,m)=(1,-1)$ mode of spin parameter $s=5$ in a M $=1.6$ M$_\sun$, R $=2.13$ R$_\sun$, X$_c = 0.35$, $\nu_{\rm rot} = 1$ d${}^{-1}$ star model, these constraints are $B_\phi/B_r \lesssim 125$ and $B_\theta/B_r \lesssim 55$ near the bottom of the radiative zone. If the horizontal field contributions were not negligible, a different frequency dependence of the magnetic shift is expected and this property can be used to identify a situation where the radial field is not dominant. 
Indeed, as argued by \citet{LD22} in the case of slowly rotating stars, the magnetic shift varies as $\propto 1/\omega_0$ for purely azimuthal fields, a frequency dependence which can be distinguished from the  $\propto 1/\omega_0^3$ dependence produced by radial fields. The numerical results of \citet{DM22} obtained for a rapidly rotating star with azimuthal fields also indicate that the increase of the magnetic shift with decreasing $\omega_0$ is much slower than $\sim 1/\omega_0^3$ (see their figure 11). 

We also use the TAR to describe the effects of the star rotation on g modes. The TAR has been tested against two oscillation codes, TOP \citep{R06,Reese09} and ACOR \citep{OD12}, which include a full treatment of the Coriolis and centrifugal accelerations. It was found the TAR provides generally a good approximation when the centrifugal deformation is not too strong \citep{BL12,OS17} although it fails in some specific frequency intervals where the so-called Rosette modes \citep{BL12,TS13} are observed or where gravito-inertial modes resonantly interact with convective core inertial modes \citep{OL20}.  

We recall that, when necessary, the approximations made to obtain simplified forms of the perturbative magnetic frequency shifts can be avoided by directly applying Eq.~(\ref{adjoint}) to a chosen field configuration and to unperturbed modes computed for a particular stellar model with full oscillation codes like TOP or ACOR. It may then be necessary to use the full expression of the Lorentz term $< \vxi_0, \LopL(\vxi_0) >$  derived in \citet{PM20}. 

\subsection{The limits of the perturbative methods}

Another limit of the present study is that perturbative methods only apply when the strength of the magnetic field is low enough. The frequency deviation produced by the magnetic field must be small compared with the unperturbed frequency (ie. $\omega_1 \ll \omega_0$). For the synthetic spectra studied in Sect.~\ref{sub2}, the magnetic shifts remained indeed much smaller than the unperturbed frequencies (the largest ratio is $\omega_1/\omega_0 = 0.048$ and corresponds to the strongest field and lowest frequency considered). Non-perturbative calculations are nevertheless necessary to determine the accuracy of the perturbative method. In a recent study, \cite{RO23} investigated the regime of strong non-perturbative magnetic fields in the particular case of a dipolar field aligned with the rotation rate. They show that perturbative methods can deviate significantly from the non-perturbative results when the field strength is close to the critical radial field above which gravity waves no longer propagate radially. As mentioned in the introduction, this critical radial field has been 
invoked to account for the unexpectedly low amplitude $\ell=1$ g modes observed in about 20\% red giants \citep{FC15}. In our case, a critical radial field  puts constraints on the possible amplitude of the magnetic shifts. Indeed, when a mode is observed, it means the internal radial magnetic field is everywhere smaller than the critical field which in turn limits the amplitude of the magnetic frequency shift. For a full observed spectrum, the condition must hold for the smallest frequency because $B_c \propto \omega^2$.  To construct the $(\ell,m)=(1,-1)$ frequency patterns of Sect.~\ref{sub2}, we did choose radial fields such that $B_r < B_c(\nu_{\rm min})$. The exact value of $B_c$, which is the subject of ongoing research \citep{RF23, RO23}, should depend on the mode and the field geometries. Using $B_c = a\sqrt{4 \pi \rho} (r/N) \omega_0^2$ with $a\approx 1/2$ as in
\citet{RF23} for $\ell=1, m=-1$ modes, we found $B_c =  24.3$ kG at the bottom of the wave cavity near $r=r_{cc}$ and $B_c =  21.6$ kG at $r=0.5 R$ for the stellar model considered. Both constraints limit the amplitude of the magnetic shifts. Finally, we note that an upper limit of $\nu_1 = \nu_0/(2 a^2)$ is obtained by assuming $B_r  = B_c(\nu_0)$ everywhere in the wave cavity and using Eq.~(\ref{two_bis}). But it is very unlikely to be reached as there is no physical reason for the actual stellar radial field to be equal to the field that prevents gravity wave propagation.

\subsection{Origin and detectability of magnetic fields in $\gamma$ Dor radiative zones}

We now discuss the possible origin of magnetic fields in $\gamma$ Dor radiative zones and whether they can produce detectable magnetic frequency shifts. 
Dynamo action generates magnetic fields in the convective core of $\gamma$ Dors. Field strengths ranging between $10$ and $100$ kG are expected from numerical simulations of a 2 M$_\sun$ convective core by \citet{BB05} and equipartition fields determined using the mixing length theory \citep{CF16}. Numerical simulations also show that the dynamo field pervades the radiative layers just above the convective core although its strength drops off sharply  \citep{BB05, AB16}. In \citet{BB05}, the field intensity decreases by approximately one order of magnitude between $r_{cc}$ and $r_{cc}+0.02R$.
A $100$ kG radial field at $r_{cc}$ with such a decrease into the overlying radiative layers produces magnetic shifts of $\nu_1 =154$ nHz for a $(\ell=1,m=-1)$ frequency corresponding to $s=$5 and the stellar model M $=1.6$ M$_\sun$, R $=2.13$ R$_\sun$, X$_c = 0.35$ previously considered. However this field exceeds the critical field which is $24.3$ kG at $r=r_{cc}+0.003R$ for the smallest frequency of a typical pattern $n \in [35,65]$. The fact that core dynamo fields can exceed the critical field for gravity wave propagation in $\gamma$ Dor stars has already been put forward in \citet{CF16} (see their figure 8 comparing $B_c$ to a core dynamo field). These authors even suggest that this mode suppression process explain why some stars in the $\gamma$ Dors or SPB instability strips are not pulsating \citep[see also][]{LB22}. If instead we consider as in Sect.~\ref{sub2} a convective core field no larger than the critical field with the same exponentially decrease, the magnetic shifts do not exceed $\sim 10 $\,nHz. They would be very difficult if not impossible to detect even with \emph{Kepler} data. However, the assumption of a sharp exponential decrease of the field strength does not take into account the decrease of convective core mass along the main sequence that leaves behind magnetic fields in radiative layers that were previously convective \citep{CF16} (see their figure 6). This process should increase the amplitude of the magnetic shifts in evolved $\gamma$ Dors. 

Other types of magnetic fields could be present in radiative interiors although we generally lack theoretical constraints on their strength and topology. This includes fossil magnetic fields possibly inherited from earlier evolutionary stages, either from the fully convective pre-main-sequence phase or from the stellar formation \citep{Braithwaite04,L14}. Such field could interact with the convective core as investigated by \citet{FB09}. Moreover, in differentially rotating radiative zones, magnetic fields could be generated by dynamo processes triggered by the Tayler instability \citep{PM23} or the Magneto-Rotational-Instability \citep{GH17}. 

Spectroscopic observations provide constraints on surface fields. The Ap/Bp stars, a class of intermediate-mass main-sequence stars, are known to harbour surface fields with typical value of
1\,kG. The radial weight function plotted in Fig.~\ref{FKr} shows that envelope fields of such strength would produce frequency shift of the same order as a $100$-kG field at the bottom of the radiative zone. While interesting this situation should remain unusual among $\gamma$ Dors because the incidence of Ap/Bp stars is very low (less than  $1$\%) in the mass range $1.4 M_\sun< M< 1.8M_\sun$  \citep{SW19}. It is more likely to find Ap/Bp stars among SPB stars, the B star HD 43317 being an example of a star with strong surface field and g-mode oscillations \citep{LB22}.  Deep spectropolarimetric surveys also revealed that much weaker magnetic fields $\sim 1$\,G might be present at the surface of many intermediate-mass stars \citep{LP09,BP16}. The magnetic shift produced by such envelope fields would be very small though. 

In Sect.~\ref{sub2}, we proposed methods to detect magnetic signatures based on the analysis of magnetically shifted asymptotic spectra.
While promising, these methods should be tested with more realistic unperturbed spectra, that is, non-asymptotic spectra that include the effects of chemical composition gradients, differential rotation, centrifugal deformation. In particular, robust generic properties of the small difference $\delta K_0 = \nu_0(1,-1,n) - \nu_0(2,-2,n)/2 $ in non-magnetic stars would be useful to detect magnetic fields. We can already notice from the study by \citet{Miglio08}
that steep chemical composition gradients seem to have a very small influence on the relation between $(\ell= 1,n)$ and $(\ell = 2,n)$ frequencies in slowly rotating stars (see their figure 27). It is also known that a weak radial differential rotation has the same effect on all high-$n$ $(\ell,m)$ modes \citep{ VM18,TO20} so that it will not modify the difference. While latitudinal differential rotation has been taken into account in studying gravito-inertial propagating waves \citep[e.g.][]{OL04,PM18}, the
effect of a latitudinal differential rotation on the frequencies of gravito-inertial modes has not been determined yet.
However, as the $(\ell,m) = (1,-1,n)$ and $ (2,-2,n)$ modes have very similar latitudinal eigenfunctions (see Fig.~\ref{Kth_compare}), we anticipate that the effects of a weak latitudinal differential rotation will be similar for the two modes. Strong radial or latitudinal differential rotation would have stronger effects (as shown in \citealt{CB18} and \citealt{TO20}, a strong radial differential rotation induces mode avoided crossings that produce large and localized deviations in the frequency pattern) but they have not yet been detected in $\gamma$ Dor stars.
The methods we propose offer new possibilities to search for internal magnetic fields in existing \emph{Kepler} or TESS data or in future observations, especially of the PLATO mission \citep{Rauer14_PLATO}, which will observe $\gamma$ Dor stars as benchmark stars.

\begin{acknowledgements}
   We acknowledge support 
   from the project BEAMING ANR-18-CE31-0001 of the French National
Research Agency (ANR) and 
from the Centre National d’Etudes Spatiales
(CNES). 
\end{acknowledgements}

\bibliographystyle{aa.bst} % style aa.bst
\bibliography{hdr}

\appendix
%\onecolumn 

\section{Simplification of the magnetic frequency shift for low frequency $\omega_0/N \ll 1$ gravito-inertial modes}\label{approx}

The expression of the magnetic shift can be significantly simplified for low frequency gravito-inertial modes. Such a simplification has been proposed by \citet{HZ05} for low frequency g modes. 
Here, we determine the conditions under which it holds for gravito-inertial modes. First, in the low frequency regime, the horizontal displacement $\xi_h$ dominates the radial one $\xi_r$. Indeed, from mass conservation, we have $\xi_h/\xi_r \sim  k_r/k_h$ where $k_r$ and $k_h$ are radial and horizontal wave vectors associated with the eigenfunction. From the WKB analysis of the radial eigenvalue problem we have $k_r = \frac{N}{\omega_0} \frac{\sqrt{\Lambda}}{r}$. The azimuthal wavevector is estimated from the azimuthal number $m$ by $k_\phi \sim m/r$ while the latitudinal wave vector $k_\theta$ can be estimated from the half width of the Hough functions in the high-$s$ regime, that is $k_\theta \sim (|m|s)^{1/2}/r$ when $\Lambda \sim m^2$ and $k_\theta \sim (\sqrt{\Lambda}s)^{1/2}/r$ when $\Lambda \neq m^2$. It follows that $k_r/k_\theta \sim (N/2\Omega)^{1/2} (N/\omega_0)^{1/2}$ when $\Lambda \sim m^2$ while $k_r/k_\theta \sim (N/2\Omega)^{1/2} (N/\omega_0)^{1/2}$ for the r mode and $k_r/k_\theta \sim (N/\omega_0) $ for the axisymmetric dipolar mode. From the same estimate, we find that the radial derivatives terms  dominate (e.g. $r\partial_r\xi_h \sim rk_r\xi_h \gg \xi_h$).  
Assuming in addition that the magnetic field varies over large scales $L$, such that $Lk_r \gg 1$, the terms involving $B_r$ dominate over the $B_\phi$ terms if  $B_\phi/B_r \ll k_r/k_\phi$ and $B_\theta/B_r \ll k_r/k_\theta$. This translates into  $B_\phi/B_r \ll (N/\omega_0)$ and 
$B_\theta/B_r \ll (N/2\Omega)^{1/2} (N/\omega_0)^{1/2}$ for the mode considered here. For a $(\ell,m)=(1,-1)$ mode such that $s=5$ in one-day rotation period, M $=1.6$ M$_\sun$, R $=2.13$ R$_\sun$, X$_c = 0.35$ stellar model, the anisotropy constraints on the
magnetic field near the bottom of the radiative zone reads $B_\phi/B_r \lesssim 125$ and $B_\theta/B_r \lesssim 55$. 

As in \citep{LD22}, these assumptions allow us to simplify the Lorentz operator. After some algebra and neglecting a surface term by assuming that it is negligible with respect to the volume integral term, we get:
\begin{equation}
< \vxi_0, \LopL(\vxi_0) > = \frac{1}{\mu_0} \int_V \vec{B'}_0^* \cdot  \vec{B'}_0 \;dV
\end{equation}
where $\vec{B'}_0=\nab \wedge (\vxi_0 \wedge \vec{B})$  
takes the simplified form:
\begin{equation}
 \vec{B'}_0 = e^{i(m \phi + \omega_0 t)} \tvect{0}{\frac{1}{r} \frac{\partial}{\partial r} (r \xi_h(r) B_r) H_\theta}{\frac{i}{r} \frac{\partial}{\partial r} (r \xi_h(r) B_r) H_\phi}
\end{equation}
which leads to
\begin{equation}
 \vec{B'}_0^* \cdot  \vec{B'}_0 = \frac{1}{r^2} \left| \frac{\partial}{\partial r} \lp r \xi_h(r) \rp \right|^2 B^2_r (H_\theta^2 + H_\phi^2)
\end{equation}
and:
\begin{equation}
< \vec{\xi}_0, {\cal L}_1(\vec{\xi}_0)>= \!\!\int_{r_i}^{r_o} \! \left| \frac{\partial}{\partial r}\!\! \lp r \xi_h \rp \right|^2 \!\!\int_0^{2\pi}\!\!\!\! \int_0^\pi \!\!\frac{B^2_r}{\mu_0} \!\!\lp H_\theta^2 + H_\phi^2 \rp \sin \! \theta \; d\theta \; d \phi \; dr
\end{equation}

\noindent From  $\vec{\xi}_0^* \cdot \vec{\xi}_0 = \xi_r^2 H_r^2 + \xi_h^2 (H_\theta^2 + H_\phi^2)$, we have
\begin{equation}
< \vec{\xi}_0, \vec{\xi}_0> = 2 \pi \int_{r_i}^{r_o} \int_0^{\pi} \rho r^2 \lp \xi_r^2 H_r^2 + \xi_h^2 (H_\theta^2 + H_\phi^2) \rp \; \sin \theta \;d\theta \; dr
\end{equation}
where the first term $\xi_r^2 H_r^2$ can be neglected for low frequency gravito-inertial modes.

In the TAR approximation,  the Coriolis term $2 i \vec{\Omega} \wedge \vec{\xi}_0$ simplifies into $2 i \Omega \cos \theta \;\er \wedge \vec{\xi}_0$,  from which we get:
\begin{equation}
< \vec{\xi}_0, 2 i \vec{\Omega} \wedge \vec{\xi}_0> = 8 \pi \Omega \int_{r_i}^{r_o} \int_0^{\pi} \rho r^2 \xi_h^2 \; dr \int_0^\pi H_\theta H_\phi \cos \theta \sin \theta \; d\theta 
\end{equation}

Then, for high-order TAR gravito-inertial modes, the magnetic frequency shift reads

\begin{equation}
\omega_1 = \frac{\int_{r_i}^{r_o} | \frac{\partial}{\partial r} \lp r \xi_h(r) \rp |^2 \int_0^{\pi} \int_0^{2\pi} B^2_r(r,\theta,\phi) \; d \phi \lp H_\theta^2 + H_\phi^2 \rp \sin \theta \; d\theta \; dr}{4 \pi \mu_0 \omega_0 \lc \int_{r_i}^{r_o} \rho r^2 \xi_h^2 \; dr\rc \lc \int_0^{\pi} \lp H_\theta^2 + H_\phi^2 - \frac{2 \Omega}{\omega_0} H_\theta H_\phi \cos \theta \rp \sin \theta \; d\theta \rc}
\end{equation}

Then by introducing $\langle B_r^2 \rangle_\phi (r,\theta) = \frac{1}{2\pi} \int_0^{2\pi} B_r^2(r,\theta,\phi) d\phi$ in this equation, we obtain Eq.~(\ref{tar}). 

\section{Computations of $\Lambda$ and the Hough functions}\label{aa}

Following \citet{WB16}, we used a numerical method based on 
Legendre polynomial 
expansion to compute first the eigenvalue $\Lambda$ and the eigenfunction $H_r$ of the Laplace's tidal equation~(\ref{Laplace}) and then the horizontal Hough functions defined by:
\begin{equation}\label{ht}
 H_\theta = \frac{1}{(1-\mu^2s^2)\sqrt{1-\mu^2}} \left( -(1-\mu^2)\frac{d H_r}{d\mu} + ms\mu H_r \right)   
\end{equation}
and
\begin{equation}\label{hp}
H_\phi  = \frac{1}{(1-\mu^2s^2)\sqrt{1-\mu^2}} \left( -s\mu(1-\mu^2)\frac{d H_r}{d\mu} + m H_r \right).
\end{equation}  

The integrals involved in the $F$ and $T_{\!\!B}$ terms are then calculated using Gauss-Legendre quadrature.
%mention error in Wang ?

\section{Asymptotic form of the $F$ and $T_{\!\!B}$ factors}\label{bb}

The Hough functions and $\Lambda$ can be expressed in closed form in the $s \gg1$ limit \citet{T03}. In this Appendix, we present these solutions for the modes most frequently observed in $\gamma$ Dors (Sect.~\ref{present}), and use them to derive asymptotic forms of $F$ (Sect.~\ref{F}) and $T_{\!\!B}$  (Sect.~\ref{T}), two terms involved in the expression of the magnetic shift.

\subsection{Asymptotic form of $\Lambda$, $H_\theta$, $H_\phi$}
\label{present}
The asymptotic solutions depend on whether $\Lambda \neq m^2$ or $\Lambda \approx m^2$.

\subsubsection{$\Lambda \neq m^2$}

This concerns the $(\ell=1, m=0)$ g mode and the $(k=-2,m=1)$ r mode. The asymptotic $\Lambda$ is given by:

\begin{equation}
\frac{m s -m^2 + \Lambda}{\sqrt{\Lambda} s}= 2 p + 1
\end{equation}
where $p$ is an integer and the horizontal Hough functions are:

\begin{align}
\sin \theta  H_\theta &  = \hat{\Theta} = H_p(u) e^{-u^2/2} \\
\sin \theta  H_\phi & = -\tilde{\Theta} = - m \frac{\lp \Lambda^{1/2} s \rp^{1/2}}{m^2 -\Lambda} \left[ p (1+\frac{\sqrt{\Lambda}}{m}) H_{p-1}(u) \right. \nonumber\\
& \left. + \frac{1}{2} (-1 + \frac{\sqrt{\Lambda}}{m}) H_{p+1}(u) \right] e^{-u^2/2}
\end{align}
where $u=\lp \Lambda^{1/2} s \rp^{1/2} \mu$, with $\mu=\cos \theta$ and
$H_p$ is the "physical" Hermite polynomial of order $p$, and 
$\hat{\Theta}$ and $-\tilde{\Theta}$ are the notations used by \citet{T03}.
 
For the axisymmetric mode  $(\ell=1, m=0)$ g mode, we get 
\begin{align}
p & = 0 \\
\Lambda &= s^2 \\
u & = \mu s \\
\sin \theta  H_\theta & = e^{-u^2/2} \\ 
\sin \theta  H_\phi & = u e^{-u^2/2}
\end{align}
and for the $(k=-2, m=1)$ r mode
\begin{align}
p & = 1 \\
\Lambda &= \frac{1}{9} \lp 1 - \frac{8}{9s} \rp^2 \\
u & = \mu \sqrt{\sqrt{\Lambda}s} \\
\sin \theta  H_\theta & = 2 u e^{-u^2/2} \\
\sin \theta  H_\phi & = - \frac{\sqrt{\sqrt{\Lambda}s}}{1-\Lambda} \lp (1+\sqrt{\Lambda}) -\frac{1}{2} (1-\sqrt{\Lambda})(4u^2-2)\rp  e^{-u^2/2}
\end{align}
The value of $\Lambda$ comes from \citet{T20} that improved the expression given in \cite{T03}. In the following we shall simply use $\Lambda = 1/9$ as we find it is better approximation to the numerical solution of $\Lambda$.

\subsubsection{$\Lambda \approx m^2$}

The sectoral prograde modes $(\ell,m=-\ell)$ verify
\begin{align}
\Lambda & =m^2 \frac{2ms}{2ms+1} \\
\sin \theta  H_\theta & = \hat{\Theta} = - \frac{1}{\lp - m s \rp^{1/2}} \frac{m^2}{2ms+1} \tau e^{-\tau^2/2} \\ 
\sin \theta  H_\phi & = -\tilde{\Theta} = m \lp \frac{\tau^2}{2ms+1} +1 \rp e^{-\tau^2/2}
\end{align}
where $\tau = \lp - m s \rp^{1/2} \mu$.

Thus for the dipolar prograde mode $(\ell=1, m=-1)$:
\begin{align}
\Lambda &= \frac{2s}{2s-1} \\
\tau & = \mu s^{1/2} \\
\sin \theta  H_\theta & = \frac{1}{s^{1/2}(2s-1)} \tau e^{-\tau^2/2} \\
\sin \theta  H_\phi & = \lp \frac{\tau^2}{2s-1} -1 \rp e^{-\tau^2/2}
\end{align}
and for the quadrupolar sectoral prograde mode $(\ell=2, m=-2)$:
\begin{align}
\Lambda &= \frac{16s}{4s-1} \\
\tau & = \mu (2s)^{1/2} \\
\sin \theta  H_\theta & = \frac{4}{(2s)^{1/2}(4s-1)} \tau e^{-\tau^2/2} \\
\sin \theta  H_\phi & = 2 \lp \frac{\tau^2}{4s-1} -1 \rp e^{-\tau^2/2}.
\end{align}

\subsection{Asymptotic form of the $F$ factors}
\label{F}

The expression (\ref{FF}) of $F$, that we recall here,
\begin{equation}
F = \Lambda \frac{\int_0^\pi \lp H_\theta^2 + H_\phi^2 \rp \sin \theta \; d\theta}{\int_0^{\pi} \lp H_\theta^2 + H_\phi^2 - \frac{2 \Omega}{\omega_0} H_\theta H_\phi \cos \theta \rp \sin \theta \; d\theta}
\end{equation}
can be written $F = \frac{\Lambda}{1-s G_2/G_1}$ with
\begin{align}
G_1 &= \int_0^\pi \lp H_\theta^2 + H_\phi^2 \rp \sin \theta \; d\theta \\
G_2 & =  \int_0^{\pi} H_\theta H_\phi \cos \theta \sin \theta \; d\theta
\end{align}

In the following we approximate $\sin^2 \theta \approx 1$ as it is the basic assumption of \citet{T03} asymptotic solutions. 
For each of the four modes considered, we rewrite the integrals $G_1$ and $G_2$ using $u$ or $\tau$ variables instead of $\theta$. We then get exact and then asymptotic $s \gg 1$ expressions of $G_1$ and $G_2$ using Wolfram alpha. The resulting asymptotic expressions for $F$ are:

\begin{align}
F_{1,1}  & \approx 1+ \frac{1}{4s}+ \frac{1}{8s^2}  \\
F_{2,-2}  & \approx 4 \lp 1 + \frac{1}{8s} + \frac{1}{32s^2} \rp \\
F_{-2,1}   & \approx \frac{1}{9} \lp 1 - \frac{8}{9} \sqrt{\frac{s}{3\pi}} e^{-s/3} \rp \approx \frac{1}{9} \\ 
F_{1,0}  & \approx \frac{3}{2} s^2 \lp 1 + \frac{2}{3 \sqrt{\pi}} s e^{-s^2} \rp \approx \frac{3}{2} s^2
\end{align}

\begin{figure}
\centering
\includegraphics[width=\hsize]{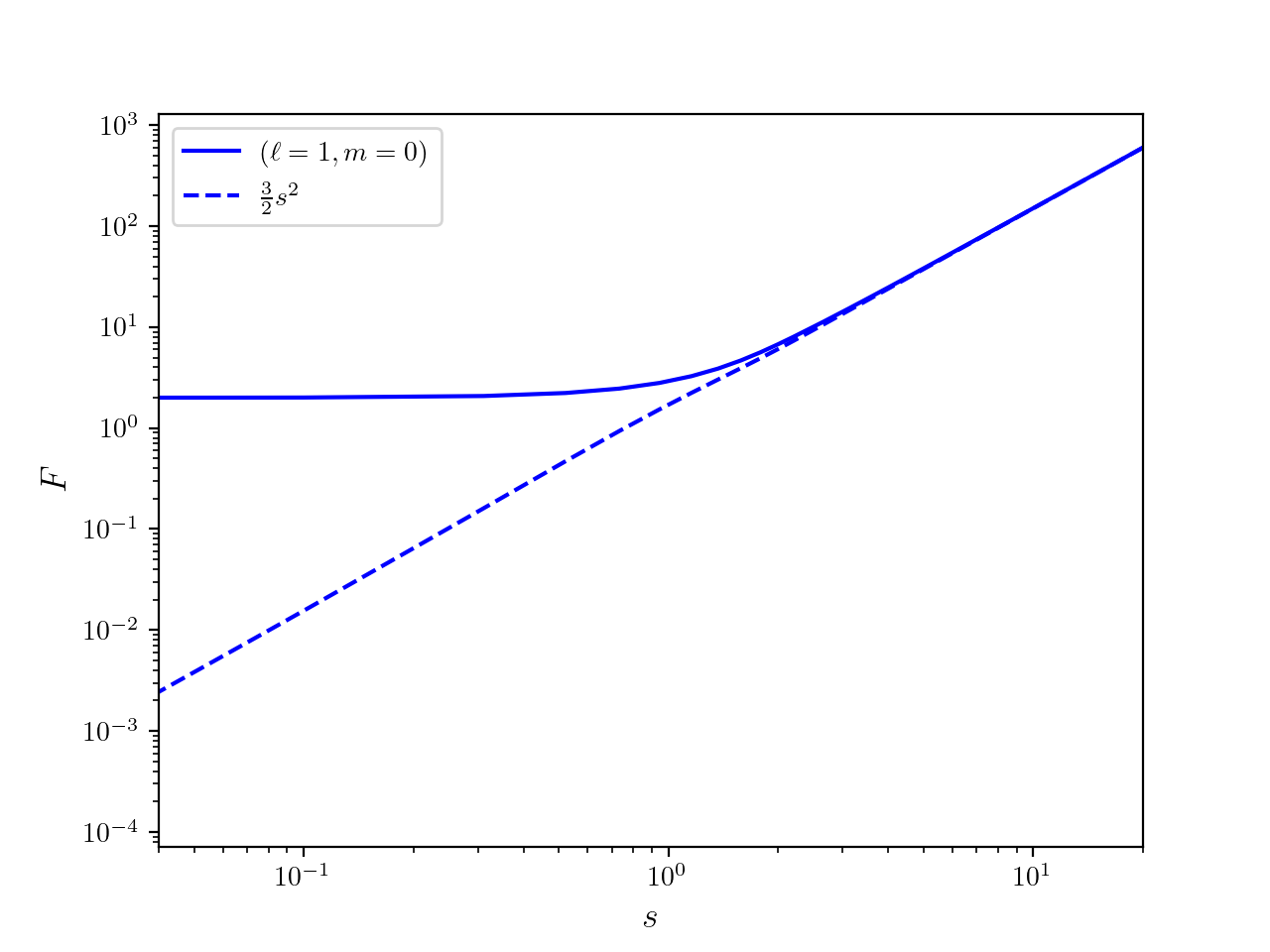}
\caption{The $F$ factor of the 
the zonal dipolar mode $\ell = 1, m=0$ (blue) as a function of the spin parameter $s$ and its asymptotic $s \gg 1$ form (dashed lines).}
         \label{other}
\end{figure}

Fig~\ref{other} presents the comparison of the asymptotic expression and numerical calculation of $F$ for the $(\ell=1,m=0$) mode.

\subsection{Asymptotic form of the $T_{\!\!B}(s)$ factor}
\label{T}

For large spin parameters $s$, the weight function $K_\theta$ is focused near the equator. Asymptotically it plays the same role as a Dirac function so that $T_{\!\!B}(s) \rightarrow \langle B_r^2 \rangle_{\phi,r}/ \langle B_r^2 \rangle$ as $s \rightarrow +\infty$.
In order study the next order in the high-$s$ asymptotic regime, we expand $\langle B_r^2 \rangle_{\phi,r}$ around the equator using the $\mu=\cos\theta$ variable, namely:
\begin{align}
\langle B_r^2 \rangle_{\phi,r}(\mu) &= \langle B_r^2 \rangle_{\phi,r}(\mu=0) + \mu \frac{d \langle B_r^2 \rangle_{\phi,r}}{d \mu}(\mu=0) \nonumber \\
& + \frac{\mu^2}{2} \frac{d^2 \langle B_r^2 \rangle_{\phi,r}}{d \mu^2}(\mu=0) + ...
\end{align}
\noindent From the definition of the factor $T_{\!\!B}(s)$, Eq.~(\ref{TT}), we have:
\begin{align}
T_{\!\!B}(s) \langle B_r^2 \rangle = & \int_0^\pi K_\theta(\theta) \langle B_r^2 \rangle_{\phi,r} \sin \theta \; d\theta \\
= & \langle B_r^2 \rangle_{\phi,r}(\mu=0) + \left(\int_{-1}^{1} \mu K_\theta(\mu) \; d\mu \right) \frac{d \langle B_r^2 \rangle_{\phi,r}}{d \mu}(\mu=0) \nonumber \\
& + \frac{1}{2} \left(\int_{-1}^{1} \mu^2 K_\theta(\mu) \; d\mu \right) \frac{d^2 \langle B_r^2 \rangle_{\phi,r}}{d \mu^2}(\mu=0) + ... \\
= & \langle B_r^2 \rangle_{\phi,r}(\mu=0) \nonumber \\
&+ \frac{1}{2} \left(\int_{-1}^{1} \mu^2 K_\theta(\mu) \; d\mu \right) \frac{d^2 \langle B_r^2 \rangle_{\phi,r}}{d \mu^2}(\mu=0) + ...\\
= & \langle B_r^2 \rangle_{\phi,r}(\theta= \pi/2) \nonumber \\
&+ \frac{1}{2} \left(\int_{-1}^{1} \mu^2 K_\theta(\mu) \; d\mu \right) \frac{d^2 \langle B_r^2 \rangle_{\phi,r}}{d \theta^2}(\theta=\pi/2) + ...
\end{align}
where we used that the integral associated with the first derivative term vanishes because the weight function $K_\theta$ is symmetric with respect to the equator and
that, at the equator, the second derivative with respect to the $\mu$ variable relates simply to the $\theta$ one.

We expect that this expression is approximately valid as long as the variations of $\langle B_r^2 \rangle_{\phi,r}$ around the equator 
arise on lengthscales larger than the latitudinal width of the weight function. This will depend on both the spin parameter and the magnetic field configuration. At very high $s$,
$\langle B_r^2 \rangle_{\phi,r}$ should vary on very small scales for this approximation to fail. Conversely, at small $s$, say $s \sim 2$, the approximation would need $\langle B_r^2 \rangle_{\phi,r}$ to be nearly uniform to be accurate.

Below, the integral involving the second derivative term is evaluated at high $s$ for the four modes of interest. We thus obtain

\begin{align}
T_{\!\!B}^{1,-1}(s) \langle B_r^2 \rangle &= \langle B_r^2 \rangle_{\phi,r}(\theta=\pi/2) + \frac{1}{4 s} \frac{d^2 \langle B_r^2 \rangle_{\phi,r}}{d \theta^2}(\theta=\pi/2) \\
T_{\!\!B}^{2,-2}(s) \langle B_r^2 \rangle &= \langle B_r^2 \rangle_{\phi,r}(\theta=\pi/2) + \frac{1}{8 s} \frac{d^2 \langle B_r^2 \rangle_{\phi,r}}{d \theta^2}(\theta=\pi/2)\\
T_{\!\!B}^{-2,1}(s) \langle B_r^2 \rangle &= \langle B_r^2 \rangle_{\phi,r}(\theta=\pi/2) + \frac{3}{4s} \frac{d^2 \langle B_r^2 \rangle_{\phi,r}}{d \theta^2}(\theta=\pi/2) \\
T_{\!\!B}^{1,0}(s) \langle B_r^2 \rangle &= \langle B_r^2 \rangle_{\phi,r}(\theta=\pi/2) + \frac{5}{12 s^2} \frac{d^2 \langle B_r^2 \rangle_{\phi,r}}{d \theta^2}(\theta=\pi/2)
\end{align}

\section{The asymptotic magnetic frequency of oblique dipolar fields}
\label{od}
From the radial component of an oblique dipolar field $B_r(r,\theta,\phi) = B_0 b(r) (\cos \theta \cos \beta + \sin \theta \cos \phi \sin \beta)$, the two averages
$\langle B_r^2 \rangle$ and $\langle B_r^2 \rangle_{\phi,r}$ read: 
\begin{align}
\langle B_r^2 \rangle &= \frac{B_0^2}{4 \pi} \lp \int_{r_i}^{r_o} K_r(r) b^2(r) \; dr \rp \nonumber\\
&\times \int_0^{\pi} \left[ \int_0^{2\pi} \left( \cos \theta \cos\beta +  \sin \theta \cos \phi \sin \beta \right)^2 \; d \phi \right] \sin \theta \; d \theta 
\end{align}

\begin{align}
\langle B_r^2 \rangle_{\phi,r} &= \frac{B_0^2}{2 \pi} \lp \int_{r_i}^{r_o} K_r(r) b^2(r) \; dr \rp\nonumber\\
&\times \int_0^{2\pi} \lp \cos \theta \cos\beta + \sin \theta \cos \phi \sin \beta \rp^2 \; d \phi 
\end{align}

Showing that
\begin{equation}
 \int_0^{\pi} \lc \int_0^{2\pi} \lp \cos \theta \cos\beta + \sin \theta \cos \phi \sin \beta \rp^2 \; d \phi \rc \sin \theta \; d \theta = \frac{4 \pi}{3}
\end{equation}
we have 
\begin{align}
\langle B_r^2 \rangle = & \frac{B_0^2}{3} \int_{r_i}^{r_o} K_r(r) b^2(r) \; dr \\
\langle B_r^2 \rangle_{\phi,r} = & B_0^2 \int_{r_i}^{r_o} K_r(r) b^2(r) \; dr \lp \cos^2 \theta \cos^2\beta +\frac{1}{2} \sin^2 \theta \sin^2 \beta \rp 
\end{align}
and
\begin{equation}
\frac{\langle B_r^2 \rangle_{\phi,r}}{\langle B_r^2 \rangle} = 3 \cos^2 \theta \cos^2 \beta + \frac{3}{2} \sin^2 \theta \sin^2 \beta 
\end{equation}
from which we deduce Eq~(\ref{Tod}) giving $T_{\!\!B}$ for an oblique dipolar field.

With the expressions of $B_{eq}^2 =  \langle B_r^2 \rangle_{\phi,r}(\theta = \pi/2)$ and $D_{eq} = \frac{d^2 \langle B_r^2 \rangle_{\phi,r}}{d \theta^2}(\theta=\pi/2)/B_{eq}^2$ :
\begin{equation}
B_{eq}^2 =  B_0^2 \lp \int_{r_i}^{r_o} K_r(r) b^2(r) \; dr\rp \frac{\sin^2 \beta}{2} = 3 \langle B_r^2 \rangle \frac{\sin^2 \beta}{2}
\end{equation}
\begin{equation}
D_{eq} =  \frac{3 \cos^2 \beta -1}{\frac{1}{2}\sin^2 \beta}, 
\end{equation}
\noindent for an oblique dipolar field,
the high-$s$ magnetic shift formulas Eq.~(\ref{two_bis}) applied to an oblique dipolar field yield :

\begin{align}
\omega_1^{1,-1} = &\frac{A}{\omega_0^3} \lp \frac{1}{2} \sin^2 \beta + \frac{\frac{1}{2} \sin^2 \beta  + 3 \cos^2 \beta -1}{4s} \rp \\
\omega_1^{2,-2} = &\frac{4A}{\omega_0^3} \lp \frac{1}{2} \sin^2 \beta + \frac{\frac{1}{2} \sin^2 \beta  + 3 \cos^2 \beta -1}{8s} \rp \\
\omega_1^{-2,1} = &\frac{A}{9 \omega_0^3} \lp \frac{1}{2} \sin^2 \beta + \frac{3( 3\cos^2 \beta -1)}{4s} \rp \\
\omega_1^{1,0} = &\frac{6 A \Omega^2}{\omega_0^5} \lp \frac{1}{2} \sin^2 \beta + \frac{5 (3 \cos^2 \beta -1)}{12s^2} \rp 
\end{align}
\noindent where $A = \frac{{\cal I} B_0^2 \int_{r_i}^{r_o} K_r(r) b^2(r) \; dr}{2 \mu_0} = \frac{3 {\cal I} \langle B_r^2 \rangle}{2 \mu_0}$.

For the $(\ell,m)=(1,-1)$ mode, the term in parenthesis read $\frac{1}{2s}$, $\frac{1}{3}(1+\frac{1}{4s})$, $\frac{1}{2}(1-\frac{1}{4s})$ for the three angles $\beta = 0^\circ,54.7356^\circ,90^\circ$, respectively. The angle $\beta = 54.7356^\circ$ is such that $\cos^2\beta =1/3$.

\section{Ratio asymptotics}
\label{cc}

In the short-radial-wavelength asymptotic regime, the product $s \sqrt{\Lambda}$ is constant for modes having the same order $n$.
With this property and the asymptotic forms of $\Lambda$ given above, the spin parameters
of the modes $(\ell=2, m=-2)$, $(\ell=1, m=0)$, $(k=-2,m=1)$, respectively denoted
$s_{2,-2}$, $s_{1,0}$,$s_{-2,1}$ can be expressed as a function of the spin parameter
of the  $(\ell=1, m=-1)$ mode, denoted $s_{1,-1}$:

\begin{equation}
\label{s_ratios}
s_{2,-2} = s_{1,-1}/2 \;\;\;\;  s_{1,0} = \sqrt{s_{1,-1} \sqrt{\Lambda_{1,-1}}} 
\;\;\;\;  s_{-2,1} = 3 s_{1,-1}\sqrt{\Lambda_{1,-1}}. 
\end{equation}

The second and third equalities simply follow from the $\Lambda$ expressions. The first equality is valid for all the sectoral prograde modes $\ell=|m|, m<0$
as 
the equation $s \sqrt{\Lambda} = c$ with $\Lambda  =m^2 \frac{2ms}{2ms+1}$ can be written
\begin{equation}
x^3 - c^2 x + c^2/2 = 0
\end{equation}
with $x=|m|s$. Thus, if $x_0$ denotes the real solution of this equation, the spin parameters $s_{2,-2}$ and $s_{1,-1}$ verify $x_0=s_{1,-1}=2s_{2,-2}$.
We note that when $s$ (and thus $c$) is large the approximate solution $x = c -1/4$ is a better approximation than the usual $x=c$ high-$s$ solution \citep[see also][]{T20}. 

\end{document}

%% file: commands.tex
\newcommand{\lp}{ \;\left(}
\newcommand{\rp}{ \right)}
\newcommand{\nab}{ \vec{\nabla} }

\newcommand{\beq}{\begin{equation}}
\newcommand{\eeq}{\end{equation}}

\newcommand{\lc}{ \left[}
\newcommand{\rc}{ \right]}
\newcommand{\greq}{\begin{equation} \begin{array}{l}}
\newcommand{\egreq}{\end{array} \end{equation}}

\newcommand{\YL}{ Y^m_\ell }

\newcommand{\YTL}{\hat{Y}^m_\ell}

\newcommand{\DYTL}{\frac{\partial \hat{Y}^m_{\ell}}{\partial \theta}}

\newcommand{\eeqn}[1]{\label{#1}\end{equation}}

\newcommand{\beqa}{\begin{eqnarray}}
\newcommand{\eeqa}{\end{eqnarray}}
\newcommand{\beqan}{\begin{eqnarray}}
\newcommand{\eeqan}[1]{\label{#1}\end{eqnarray}}

\newcommand{\tvect}[3]{ \begin{pmatrix} #1 \\ #2 \\ #3 \end{pmatrix} }